\documentclass[letterpaper, twocolumn, 10pt]{article}
\usepackage{usenix-2020-09}
\usepackage{cite}
\usepackage{amsmath,amssymb,amsfonts}
\usepackage{algorithmic}
\usepackage{graphicx}
\usepackage{textcomp}
\usepackage{xcolor}

\usepackage{xspace}
\usepackage{lipsum}
\usepackage{enumerate}
\usepackage{pifont}
\usepackage{booktabs}
\usepackage{multirow}
\usepackage{threeparttable}
\usepackage{array}
\usepackage{graphicx}
\usepackage{subfigure} 

\usepackage{bbding}
\usepackage{epigraph} 
\usepackage{balance}

\newcommand{\sys}{{\sc Revelio}\xspace}
\newcommand{\threesize}{2.3cm}

\newcommand{\theosize}{4.5cm}

\begin{document}

\date{}

\title{\Large \bf {\sc Catch You and I Can}: Revealing Source Voiceprint Against Voice Conversion
}

\normalsize{
	\author{
	\rm Jiangyi Deng$^1$, Yanjiao Chen$^1$\thanks{Corresponding author.}~, Yinan Zhong$^1$, Qianhao Miao$^1$, Xueluan Gong$^2$, Wenyuan Xu$^1$\\
	$^1$Zhejiang University, $^2$Wuhan University
    }
}

\maketitle

\begin{abstract}

Voice conversion (VC) techniques can be abused by malicious parties to transform their audios to sound like a target speaker, making it hard for a human being or a speaker verification/identification system to trace the source speaker. In this paper, we make the first attempt to restore the source voiceprint from audios synthesized by voice conversion methods with high credit. However, unveiling the features of the source speaker from a converted audio is challenging since the voice conversion operation intends to disentangle the original features and infuse the features of the target speaker. To fulfill our goal, we develop \sys, a representation learning model, which learns to effectively extract the voiceprint of the source speaker from converted audio samples. We equip \sys with a carefully-designed differential rectification algorithm to eliminate the influence of the target speaker by removing the representation component that is parallel to the voiceprint of the target speaker. We have conducted extensive experiments to evaluate the capability of \sys in restoring voiceprint from audios converted by VQVC \cite{tang2022avqvc}, VQVC+ \cite{wu2020vqvcp}, AGAIN  \cite{chen2021again}, and BNE  \cite{liu2021any}. The experiments verify that \sys is able to rebuild voiceprints that can be traced to the source speaker by speaker verification and identification systems. \sys also exhibits robust performance under inter-gender conversion, unseen languages, and telephony networks.

\end{abstract}

\section{Introduction}\label{sec:intro}
\setlength\epigraphwidth{.9\columnwidth}
\epigraph{\textit{The nets of Heaven are wide,
but nothing escapes its grasp.}}{Lao Tzu's Tao Te Ching \cite{lin2006tao}}

Voice conversion, a commonly-used speech synthesis technique, enables a person to transform their voice to sound like another person without changing the linguistic content. While the initial incentive for voice conversion could be simply novelty and curiosity, the technological breakthrough of deep learning has made voice conversion available for real-life applications, such as voice dubbing for movies \cite{zhang2019joint}, aids for the speech-impaired \cite{veaux2013towards}, voice mimicry \cite{wu2014voice} and disguise \cite{huang2021defending}. Unfortunately, when falling into the wrong hand, voice conversion may be used to carry out misdeeds. In 2020, the U.S. district court of Columbia tried a case where hackers had heisted \$40 million by calling company managers using ``deep voice'' technology to simulate the voice of the Director \cite{bankheist}. 

Given the high frequency of phone calls nowadays and the ability to retain phone recordings, it is unsurprising that voice analysis has become a key tool for criminal forensics. However, audio data processed by speech synthesis techniques such as voice conversion may compromise the integrity of voice forensics. In particular, voice conversion distorts the features of the source speaker, making it difficult, if not impossible, to identify the true speaker. Existing works mainly focus on determining whether an audio sample is genuine or fake \cite{tak2021end, ge2021raw, chen2021pindrop, yi2021half, das2021know, chen2021ur, tomilov2021stc, leon2012evaluation, wang2020deepsonar, chen2020generalization, ma2021continual, zhang2021multi, DBLP:conf/wnsp/PerrotAC05}, but cannot trace the source of a fake audio sample. {A few previous works~\cite{DBLP:journals/tifs/ZhengLSZZ21, DBLP:journals/dsp/WangWH15, kumar2022acoustic} have attempted to restore transformed voices, but they are restricted to traditional simple frequency-domain voice transformations (e.g., pitch scaling and vocal tract length normalization)~\cite{DBLP:journals/tifs/ZhengLSZZ21, DBLP:journals/dsp/WangWH15} or man-made disguises (e.g., mimicking)~\cite{kumar2022acoustic}. As far as we know, there is a lack of success in the case of VC that adopts complex learning models~\cite{DBLP:journals/tifs/ZhengLSZZ21}.}

\begin{figure}[t]
    \centering

\includegraphics[width=3.2in, trim=320 150 310 150, clip]{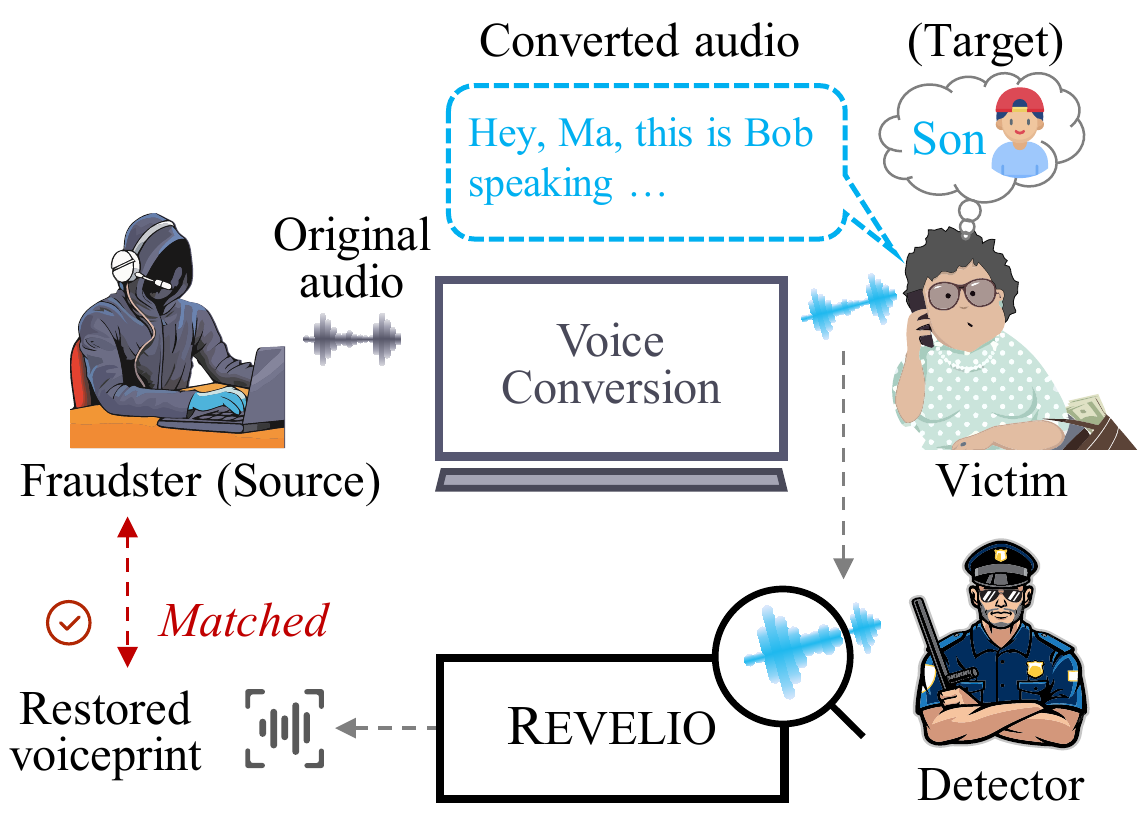}

\caption{Application scenario of \sys. A fraudster uses voice conversion to transform their voice to sound like the target person, e.g., family members or acquaintances of the victim. \sys helps restore the voiceprint of the source speaker for investigation purposes.}\label{fig:scenarios}
\end{figure}

{In this paper, we propose the first effective approach to restore the voiceprint of the source speaker from audio processed by voice conversion techniques.} Our developed system, \sys\footnote{The Revelio Charm is taught by Professor McGonagall in Transfiguration Classes in Hogwarts. The Revelio Charm reveals the true form of things or makes the invisible visible again\cite{rowling2008harry}.}, aims to re-construct the original personal features of the converted audio such that the extracted voiceprint can match the source speaker via a speaker verification or identification model. In the case of a phone scam, as shown in Figure~\ref{fig:scenarios}, \sys may assist law enforcement officials in investigations.

Nonetheless, recovering an identifiable voiceprint from voice-converted (VC) audio with high credit is a challenging task. There are two major difficulties that we need to overcome. (1) The voice conversion operation can transform the audio of a source speaker to different target speakers and the audio of different source speakers to the same target speaker, making it inaccurate to identify the source speaker based on a simple classification model. To deal with this problem, we utilize representation learning to rebuild rather than re-labeling the voiceprint of the source speaker. The key idea is for the representation learning model to learn to generate the embedding of the source speaker voiceprint given the input of a VC audio of the same speaker. To train a well-performed representation learning model, we construct a large dataset consisting of a total of {6,779,000} samples converted by four popular VC techniques with {9,691} source speakers. (2) The features contained in VC audio are a mixture of the source speaker features and the target speaker features, and, in large part, the target speaker features. To dig the hidden source speaker features buried under the target speaker features, we design a novel differential rectification algorithm. The main intuition is to remove the component of the extracted voice feature that is parallel to the feature of the target speaker and only keep the orthogonal component. In this way, the influence of the target speaker is expected to be reduced to the maximum extent.  {The voiceprint recovered by \sys is to be compared with the voiceprint pool collected by the police to determine the identity of the source speaker\footnote{{In an open-world, the voiceprint of the source speaker may not be collected, i.e., no match between the recovered voiceprint and the voiceprint in the pool (the identity of the source speaker cannot be determined) but does not affect the fact that \sys obtains the voiceprint of the source speaker. }}. Note that \sys only recovers the voiceprint of the source speaker of the audio input of the VC, but cannot guarantee that the voiceprint is organic if the input of VC is already manipulated.}.

We have conducted extensive experiments to evaluate the effectiveness of \sys in restoring voiceprint from four popular voice conversion techniques, i.e., VQVC \cite{tang2022avqvc}, VQVC+  \cite{wu2020vqvcp}, AGAIN  \cite{chen2021again}, and BNE  \cite{liu2021any}. We demonstrate that the restored voiceprint can be correctly identified as the source speaker by speaker verification or identification systems with more than 95\% accuracy (the source speaker is even unseen in the training dataset of \sys). An English-trained \sys model is shown to be able to recover voiceprint of German-, French- and Spanish-speaking VC audios. {We also show by experiments that when the VC audios are contaminated by noises from telephony codecs and 8k/4kHz subsampling (from 16kHz), the restoration capability of \sys is only slightly degraded. We study four codecs for public switched telephone network (PSTN) and voice over Internet protocol (VoIP) network in \S\ref{subsec:robustness}.}

We summarize our main contributions as follows.

\begin{itemize}
        \item {We propose the first effective approach to perform voiceprint restoration from audios processed by voice conversion techniques, which may be added as a tool of voice forensics to help trace and identify the source speaker of an audio.}

        \item We develop a novel and effective representation learning model to extract identifiable voice characteristics of the source speaker by removing the features that are highly related to the target speaker.
        \item We conduct extensive experiments to verify the effectiveness and robustness of our method, with a generated voice conversion dataset using {four} voice conversion methods featuring over {9,600} speakers.

\end{itemize}


\begin{figure}[t]
    \centering

\includegraphics[width=3.4in, trim=110 110 110 110, clip]{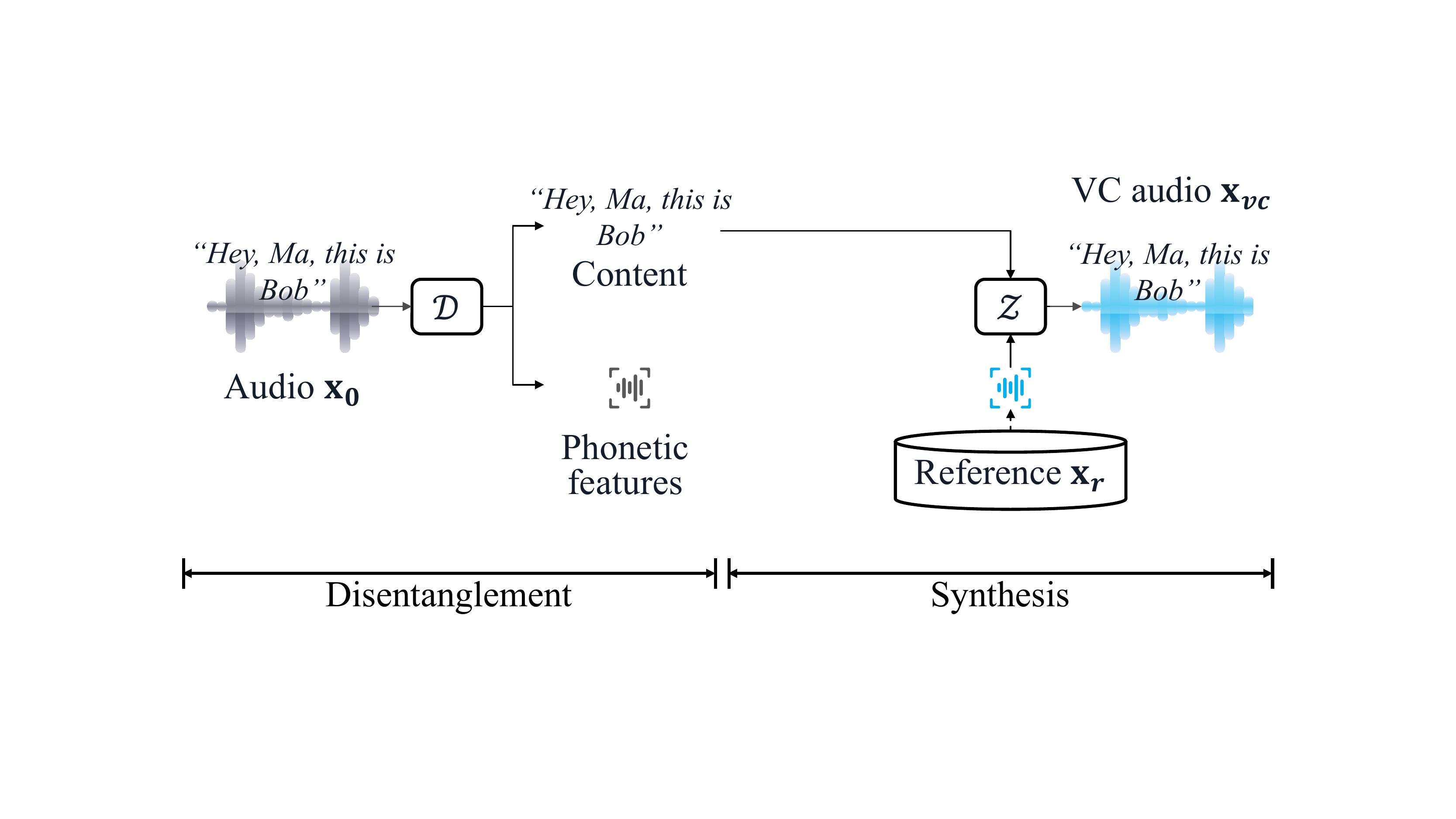}

\caption{Illustration of the process of voice conversion. The original audio sample $\mathbf{x}$ is disentangled into the content and the phonetic features. Then the original content and the phonetic features of the target speaker are synthesized into a new speech.}\label{fig:VC}
\end{figure}

\begin{figure}[t]
    \centering
\setlength{\abovecaptionskip}{0pt}
\setlength{\belowcaptionskip}{0cm}

\includegraphics[width=2.4in, trim=30 0 5 33, clip]{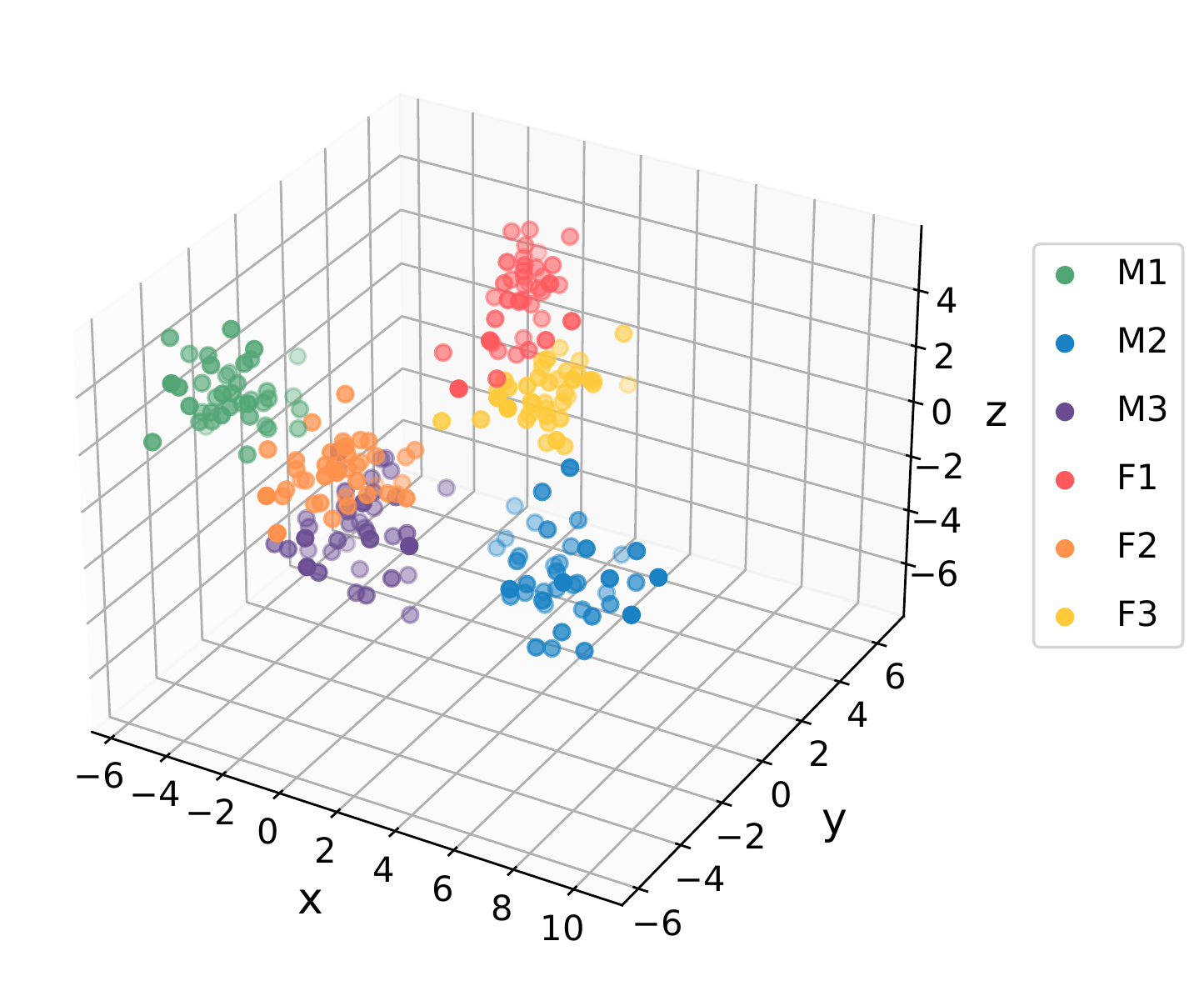}

\caption{Disentangled content samples of different speakers by BNE~\cite{liu2021any}.}\label{fig:ppgdis}
\end{figure}

\section{Background}\label{sec:background}

\subsection{Voice Conversion}

Audio data contains two kinds of information, i.e., speech contents and phonetic features.
Speech contents refer to linguistic information, i.e., ``what are the words spoken" and phonetic features refer to the way the speech contents are conveyed, i.e., ``how are the words spoken." Voice conversion (VC) alters the speaker-related phonetic features of an audio sample but maintains the linguistic information and non-speaker-related phonetic features (e.g., emotion). The objective of VC is to make human listeners mis-identify the speaker of a VC-processed audio sample as the target person (e.g., family members or friends), while the original audio sample is actually articulated by a different person (e.g., a fraudster). To achieve this goal, VC techniques attempt to replace the phonetic features of the source speaker with those of the target speaker during processing. 

As shown in Fig.~\ref{fig:VC}, given an input audio sample $\mathbf{x}_0$ that features a source speaker, the disentanglement module $\mathcal{D}$ first splits $\mathbf{x}_0$ into two components: one mostly contains the speech content, and the other mostly contains personal phonetic features of the source speaker. Then, the content component is synthesized with a set of reference audios $\mathbb{X}_\mathrm{R}=\{\mathbf{x}_r\}$ that features the target speaker in the synthesis module $\mathcal{Z}$. In this way, the converted audio $\mathbf{x}_{vc}$ sounds like the target speaker but preserves the content of $\mathbf{x}_0$.

Based on the availability of parallel reference audio samples, VC techniques can be categorized into parallel VC and non-parallel VC. Parallel VC needs to be trained with the reference sample $\mathbf{x}_r$ that has exactly the same speech content as the original sample $\mathbf{x}_0$ (i.e., the parallel corpus). Non-parallel VC can be trained with reference samples that have different speech contents from the original sample. In comparison, parallel VC is less practical than non-parallel VC due to the need for a parallel corpus for training, but the converted samples of parallel VC generally have higher quality than those of non-parallel VC also due to the benefit of the parallel corpus.  

{\textbf{Rationale of voiceprint recovery.} The imperfect disentanglement enables the recovery of the voiceprint of the source speaker. As shown in Figure~\ref{fig:VC}, the converted audio is synthesized from the content of the input audio $\mathbf{x}_0$ and the phonetic features of the target speaker. As the disentanglement module cannot perfectly separate content and phonetic features, the phonetic features of the source speaker are leaked into the content parts and synthesized into the converted audio. To demonstrate this, we collect audio samples from six volunteers (three males and three females), each speaking ``Hello World'' for 50 times. We utilize the disentanglement modules of four VCs to obtain the content of these samples, and perform a dimensional reduction via Linear Discriminant Analysis (LDA)~\cite{LDA}. We present the results in Figure~\ref{fig:ppgdis} and Figure~\ref{fig:theoretical} in the Appendix. In addition, we train a linear Support Vector Machine (SVM) to classify the disentangled content samples and achieve an accuracy of 100.0\%, 95.7\%, 88.0\%, 100.0\% for the four VCs respectively, which indicates that the content parts of different speakers are almost linearly separable. The visualization and the classification results verify that VC audios still carry information of the source speaker.
}

\subsection{Speaker Verification \& Identification}
There are two mainstream systems for examining speaker identity: speaker verification and speaker identification. The former tries to determine whether the speaker of an audio sample is the enrolled speaker or not, and the latter attempts to identify who the speaker (of an audio sample) is out of a set of enrolled speakers. To fulfill the above goals, the essential building block of both speaker verification and speaker identification systems is the voiceprint extraction method.

\textbf{Voiceprint extraction}.  Given an audio sample $\mathbf{x}\in\mathbb{R}^{1\times t}$ of an arbitrary length of $t$, a voiceprint extractor $\mathcal{V}: \mathbb{R}^{1\times t}\rightarrow \mathbb{R}^{1\times v},  \forall t \in (0, +\infty)$ maps $\mathbf{x}$ into a fixed-length voiceprint vector, representing the speaker characteristics of the audio. There are mainly two ways to build the voiceprint extractor, i.e., statistical models and Deep Neural Network (DNN)-based models. Gaussian mixture model (GMM) is a traditional statistical model to extract ivector voiceprints. ivector-PLDA~\cite{povey2011kaldi} is a popular speaker verification/identification implementation that matches ivector voiceprints via probabilistic linear discriminant analysis (PLDA). X-vector~\cite{snyder2018x}, Deep Speaker~\cite{li2017deepspeaker}, ECAPA-TDNN \cite{desplanques2020ecapa} are DNN-based voiceprint extractors, which outperform GMM as DNNs are more effective in extracting feature representations from large-scale voice datasets.
Among them, ECAPA-TDNN is the state-of-the-art voiceprint extractor based on time delay neural network (TDNN). ECAPA-TDNN outputs a 192-dimensional voiceprint vector, i.e., $v=192$. The extracted voiceprints of legitimate users during enrolment are stored as the \emph{speaker model}, which are later compared with the voiceprint of the input audio for speaker verification or identification.

\textbf{Speaker verification}. In a speaker verification system, there is a single enrolled user with the voiceprint $\mathbf{v}$. The speaker verification system aims to verify whether the speaker of an input audio is the enrolled user (e.g., the fraudster) or not. Given an input audio $\mathbf{x}$, the similarity score between the extracted voiceprint and the enrolled voiceprint is computed as $S(\mathcal{V}(\mathbf{x}), \mathbf{v})$. If the score is higher than the threshold $\phi$, $\mathbf{x}$ is considered to be matched with the enrolled speaker.
\begin{displaymath}
\mathcal{O}(\mathbf{x})=\left\{
\begin{array}{lcl}
\textrm{matched,}       &      & \textrm{if}~\mathcal{S}(\mathcal{V}(\mathbf{x}), \mathbf{v})\geq \phi,\\
\textrm{not matched,}     &      & \textrm{otherwise,}
\end{array} \right. 
\end{displaymath}
where $\mathcal{O}(\mathbf{x})$ is the output of the verification system.


\textbf{Speaker identification.} In a speaker identification system, there is a set of enrolled users with the voiceprints $\{\mathbf{v}_i\}_{i\in \mathcal{I}}$. Speaker identification aims to determine which enrolled user the speaker of the input audio is most likely to be. In particular, the voiceprint of the input audio is compared with each speaker model to calculate the similarity scores $\mathcal{S}(\mathcal{V}(\mathbf{x}), \mathbf{v}_i), i \in \mathcal{I}$. The output index of the enrolled users is the one with the highest score.
\begin{displaymath}
\mathcal{O}(\mathbf{x})=\mathop{\arg\max}\limits_{i \in \mathcal{I}}{\mathcal{S}(\mathcal{V}(\mathbf{x}), \mathbf{v}_i)},
\end{displaymath}

Without voice conversion, the source speaker of an audio sample should be correctly verified or identified by a speaker verification or identification system that has the speaker's voiceprint enrolled. Nonetheless, after voice conversion, the phonetic features of the source speaker are compromised, thus the VC-processed audio sample will evade the detection of speaker verification or identification systems.

This inspires us to seek a way to recover the voiceprint of the source speaker, which is helpful in forensic investigations. We will demonstrate the effectiveness of \sys in restoring voiceprints for both speaker verification and speaker identification tasks of different forensic investigation scenarios in $\S$\ref{sec:evaluation}.

\subsection{System Model}
Without loss of generality, we refer to the party who uses voice conversion to convert their voiceprint to a target speaker as the \emph{Dodger} and the party who wants to restore the original voiceprint as the \emph{Detector}. 

\subsubsection{Dodger Model}

We assume that the dodger has a corpus of reference audios $\mathbb{X}_\mathrm{R}=\{\mathbf{x}_r\}$ of the target speaker whose real voiceprint is $\mathbf{v}_r$, based on which the dodger uses voice conversion techniques to convert their original audio sample $\mathbf{x}_0$ into $\mathbf{x}_{vc}$ (referred to as the VC audio) that sounds like the target speaker. The voiceprint of the dodger is denoted as $\textbf{v}_d$. {Note that we assume that the original audio is uttered by the dodger. The case where the original audio is generated by text-to-speech (TTS) techniques is out of our scope since TTS leaves no source voiceprint to be recovered. 
}

\subsubsection{Detector Model}

We assume that the detector obtains the VC audio $\mathbf{x}_{vc}$ and is aware that $\mathbf{x}_{vc}$ is processed by voice conversion techniques but may not know the particular VC method being used. The detector knows the target speaker by listening to the VC audio and may obtain an audio sample of the target speaker (referred to as the evidence audio $\mathbf{x}_e$) during investigations. The evidence audio $\mathbf{x}_e$ can be non-parallel with the VC audio $\mathbf{x}_{vc}$. Having collected the VC audio $\mathbf{x}_{vc}$ and the evidence audio $\mathbf{x}_e$, the detector attempts to recover the voiceprint of the source speaker of $\mathbf{x}_{vc}$ that can be verified by a speaker verification model (e.g., the detector has a single suspect) or be identified by a speaker identification model (e.g, the detector has a group of suspects).

\begin{figure*}[t]
    \centering
\setlength{\abovecaptionskip}{0pt}
\setlength{\belowcaptionskip}{0cm}

\includegraphics[width=7in, trim=70 135 70 135, clip]{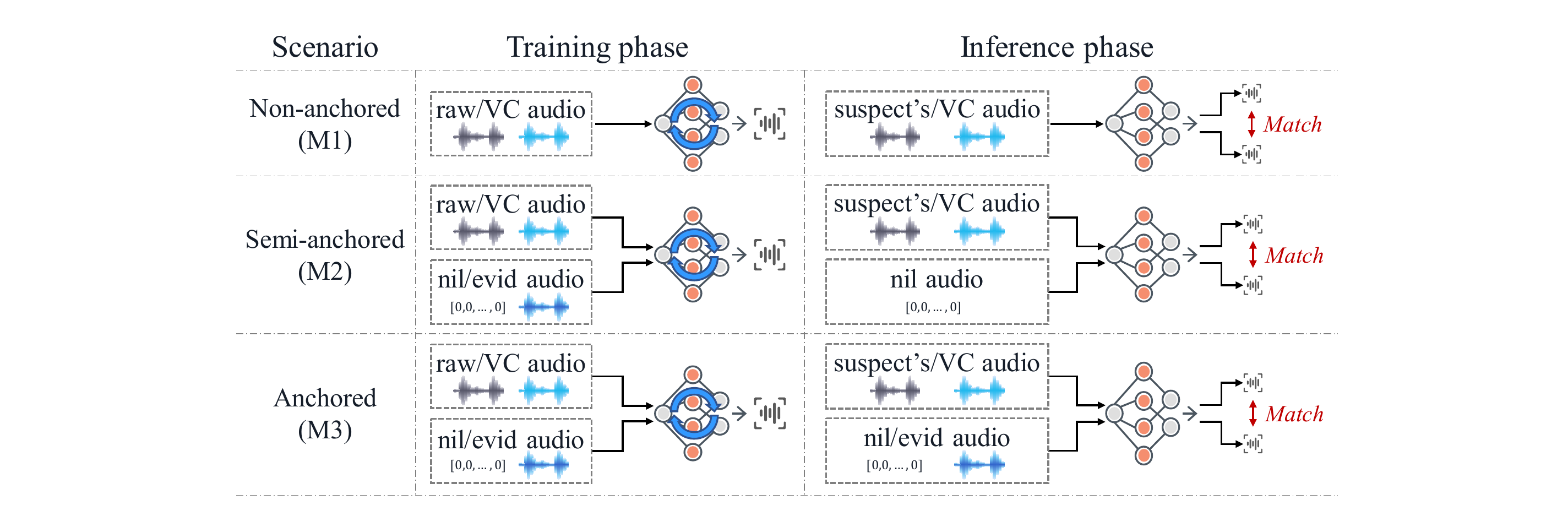}

\caption{Three materializations of \sys. In the non-anchored case M1, the evidence audios of the target speaker are unavailable. In the semi-anchored case M2, the evidence audios of the target speaker are available during the training phase, but unavailable during the inference phase. In the anchored case M3, evidence audios of the target speaker are available during the training and the inference phases. {In the semi-anchored and the anchored cases, the nil audio $\mathbf{0}$ is used as a placeholder when the evidence audio is unavailable.}}\label{fig:threemodel}
\end{figure*}

\section{Problem Formulation \& Materialization}
In this section, we first formulate the voiceprint restoration as a representation learning problem. Then we materialize the representation learning problem in three different scenarios.

\subsection{Representation Learning Problem Formulation}\label{subsec:formulation}
Given the raw audio $\mathbf{x}_0=[x_1, \cdots, x_t]\in \mathbb{R}^{1\times t}$ of length $t$, the VC audio $\mathbf{x}_{vc}$ is generated via voice conversion techniques to mimic the target speaker.  We have $\mathbf{x}_{vc} = \psi(\mathbf{x}_0, \mathbb{X}_\mathrm{R})$, where $\psi(\cdot, \cdot)$ denotes the conversion function of the VC technique and $\mathbb{X}_\mathrm{R}=\{\mathbf{x}_r\}$ contains reference samples of the target speaker. As a result of voice conversion, the similarity score between $\mathcal{V}(\mathbf{x}_{vc})$ and $\mathbf{v}_r$ greatly increases,  i.e., $\mathcal{S}(\mathcal{V}(\mathbf{x}_{vc}), \mathbf{v}_r) \gg \mathcal{S}(\mathcal{V}(\mathbf{x}_0), \mathbf{v}_r)$, and the similarity score between $\mathcal{V}(\mathbf{x}_{vc})$ and  $\mathbf{v}_d$ drops to a great extent, i.e., $\mathcal{S}(\mathcal{V}(\mathbf{x}_{vc}), \mathbf{v}_d) \ll \mathcal{S}(\mathcal{V}(\mathbf{x}_0), \mathbf{v}_d)$, making it difficult (if not impossible) for the detector to unveil the identity of the dodger. 

The objective of \sys is to restore $\mathbf{v}_d$ from $\mathbf{x}_{vc}$ with high probability, which can be formulated as a representation learning problem $\mathcal{E}: \mathbb{R}^{1\times t}\rightarrow \mathbb{R}^{1\times v}, \forall t \in (0, +\infty)$. The representation learning model $\mathcal{E}$ aims at minimizing the distance between the output vector (i.e., the reconstructed voiceprint) and the voiceprint of the dodger.
\begin{equation}\label{equ:obj2}
\max \ \mathcal{S}\big(\mathcal{E}(\mathbf{x}_{vc}, *|\theta), \mathbf{v}_d\big),
\end{equation}
where $\theta$ is the parameters of the representation learning model $\mathcal{E}$ and $*$ denotes any extra information available to the detector, e.g., an evidence audio $\mathbf{x}_e$ from the target speaker.

To train the representation learning model $\mathcal{E}$, we resort to the final objective of identifying the dodger. More specifically, we integrate $\mathcal{E}$ into a classification model that aims to classify the reconstructed voiceprint (i.e., the output of $\mathcal{E}$) into the label that represents the dodger. In other words, $\mathcal{E}$ performs feature extraction for the classification task, followed by which a fully-connected layer is attached to produce the final classification result. We denote the classification model as $\mathcal{C} = [\mathcal{E} \oplus \mathcal{F}]$, where $\mathcal{F}$ is the final layer.

The loss function of the classification model is defined as the cross entropy loss w.r.t the model parameter $\theta$.
\begin{equation}\label{equ:obj1}
\begin{aligned}
& \mathcal{L}(\theta) \triangleq -\sum_{i}{y_i\log(p_i)}, \\
\end{aligned}
\end{equation}
where $p_i = \mathcal{C}(\mathbf{x}_{vc}, *|\theta)$ is the output confidence score of the classification model regarding the $i$-th class. $y_i = 1$ if $i$ is the real dodger and $y_i=0$ otherwise. $\mathcal{F}$ usually adopts the softmax activation function, i.e., 
\begin{equation}\label{equ:softmax}
p_i= \frac{\exp(\mathcal{E}_i(\mathbf{x}_{vc}, *))}{\sum_j \exp(\mathcal{E}_j(\mathbf{x}_{vc}, *))},
\end{equation}
\noindent where $\mathcal{E}_i(\mathbf{x}_{vc}, *)$ is the $i$-th output of $\mathcal{E}$. Note that softmax does not explicitly drive the voiceprint towards the dodger voiceprint $\mathbf{v}_d$ and away from the voiceprints of other suspects. We will enhance the discriminative ability of softmax in our design in \S \ref{subsec:enhance}.

After training the classification model $\mathcal{C}$, the resulting sub-model $\mathcal{E}$ can be obtained as our voiceprint restoration model.

\begin{figure}[t]
    \centering
\setlength{\abovecaptionskip}{0pt}
\setlength{\belowcaptionskip}{0cm}

\includegraphics[width=3.in, trim=180 5 150 10, clip]{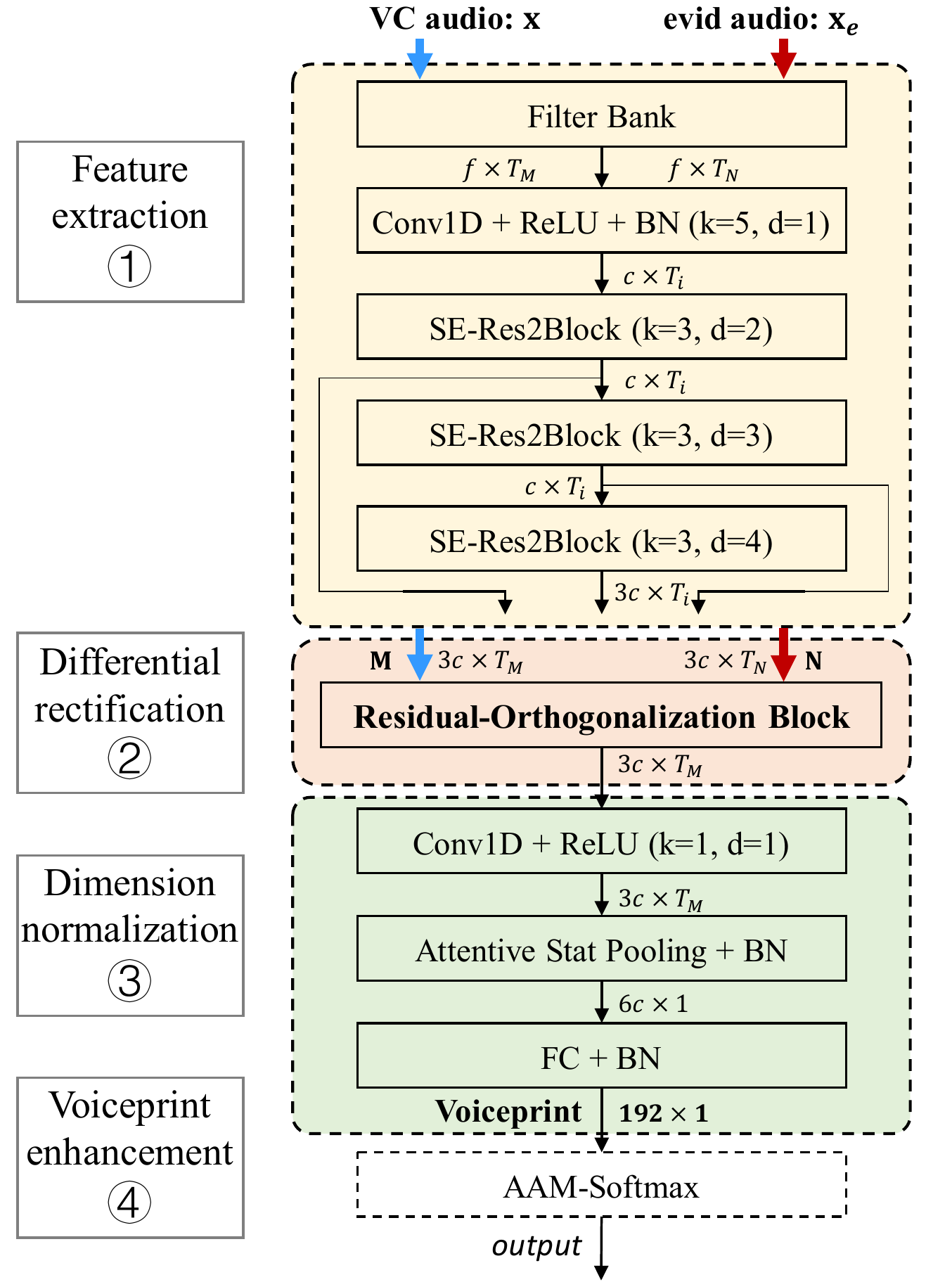}

\caption{\sys model architecture. The feature extraction block \ding{192} extracts crude voiceprint from the VC audio. The differential rectification block \ding{193} eliminates the influence of the target speaker based on the evidence audio. The dimension normalization block \ding{194} reshapes the voiceprint into a fixed-length vector. The voiceprint enhancement block \ding{195} further alters the voiceprint to be more discriminative from the voiceprints of other suspects.}\label{fig:model}
\end{figure}

\begin{figure}[tt]
    \centering
\setlength{\abovecaptionskip}{0cm}
\setlength{\belowcaptionskip}{0cm}
\subfigure[Residual Orthogonalization Block architecture]{
    \includegraphics[width=2.6in, trim=200 155 200 165, clip]{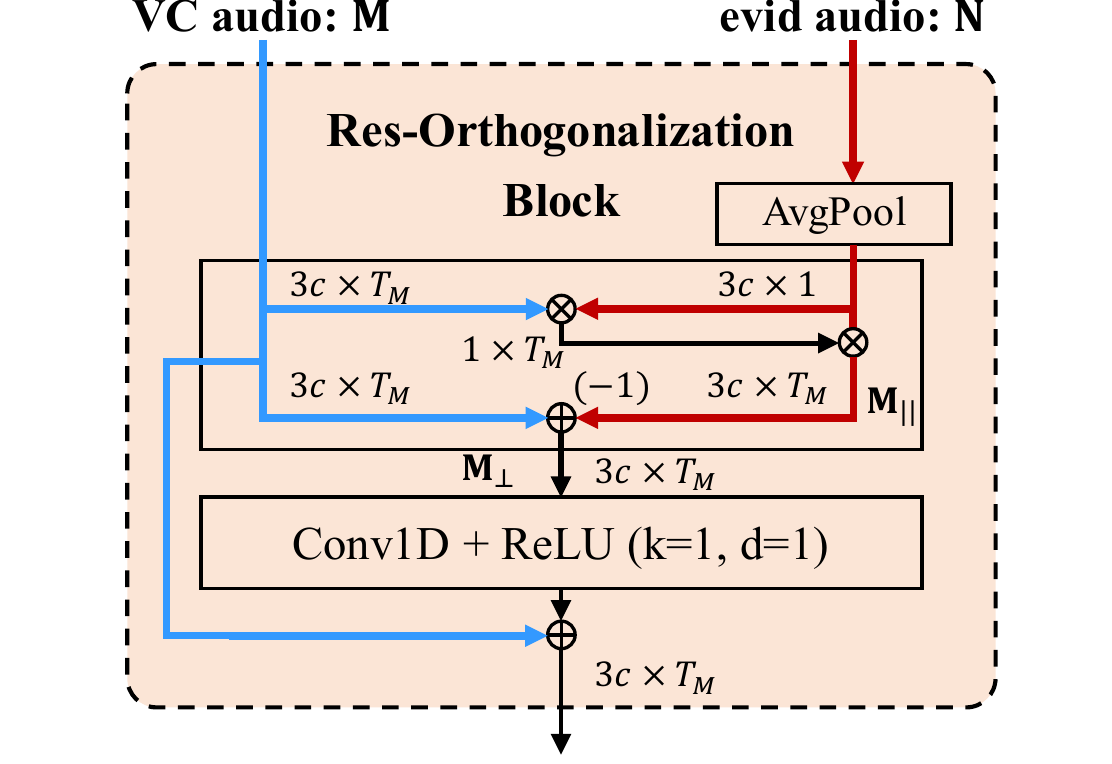}
    \label{fig:resorth1}
    }
\\
\subfigure[Extracting Invariant Features by Orthogonalization]{
    \includegraphics[width=2.2in, trim=170 170 220 140, clip]{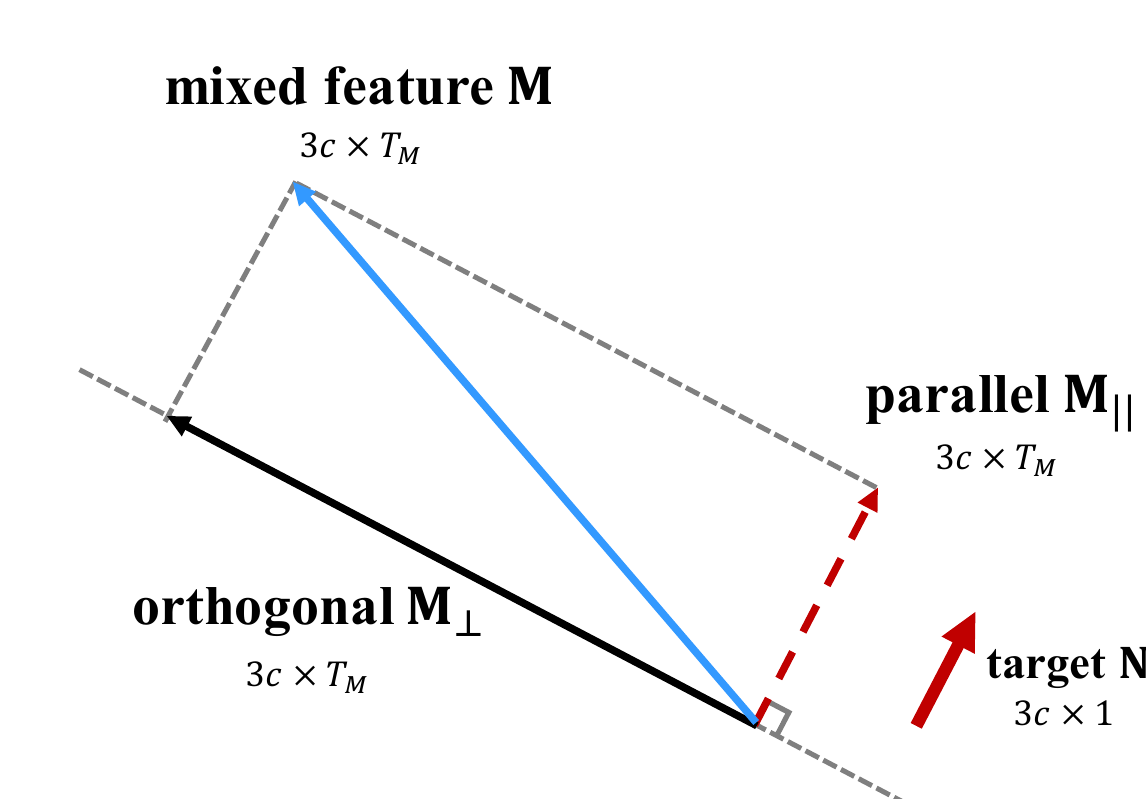}
    \label{fig:resorth2}
    }

\caption{The design of the Residual Orthogonalization Block. The block decomposes the voiceprint into two components that are parallel and orthogonal to the voiceprint of the target speaker. By subtracting the parallel component, the influence of the target speaker voiceprint is mitigated.}\label{fig:resorthblock}
\end{figure}

\subsection{Materialization Scenarios}\label{subsec:scenario}

According to the availability of the evidence audio $\mathbf{x}_e$ during the training phase and the inference phase, we materialize \sys in three cases, as shown in Figure~\ref{fig:threemodel}.

\begin{itemize}

\item \textbf{Non-anchored case}. In a non-anchored case, the evidence audio $\mathbf{x}_e$ that features the target speaker of the VC audio $\mathbf{x}_{vc}$ is unavailable, both during the training phase and the inference phase. $\mathcal{E}$ has a single input in this case. During the training phase, $\mathcal{E}$ learns to output the ground-truth voiceprint $\mathbf{v}_d$ given a training VC audio $\mathbf{x}_{vc}^{train}$ or a training raw audio $\mathbf{x}_0^{train}$ (both originates from the dodger). 

\item \textbf{Semi-anchored case}. In a semi-anchored case, the evidence audio $\mathbf{x}_e$ that features the target speaker of the VC audio $\mathbf{x}_{vc}$ is available during the training phase but unavailable during the inference phase. $\mathcal{E}$ has dual inputs in this case. During the training phase, $\mathcal{E}$ learns to output the ground-truth voiceprint $\mathbf{v}_d$ given a training VC-evidence audio pair $[\mathbf{x}_{vc}^{train}, \mathbf{x}_e^{train}]$ or a training raw-nil audio pair $[\mathbf{x}_0^{train}, \mathbf{0}]$. 

\item \textbf{Anchored case}. The anchored case M3 is the same as the semi-anchored case M2 except that during the inference phase,

\end{itemize}

The three above cases can be applied to different scenarios. 
For the non-anchored case, the detector only needs a training dataset of VC audios and their ground-truth labels of the dodger identity during training and a piece of VC audio during inference. In the semi-anchored case, the detector also needs to obtain the corresponding evidence audio for each VC audio during training.  In the anchored case, the detector further should have the evidence audio for the VC audio during inference. {Thus, in the anchored case we assume that the detector knows the input audio is processed by VC (possibly via existing VC detection techniques), and obtains an evidence audio from the target for inference. But we do not make this assumption in the non-anchored or the semi-anchored cases. Note that given a non-VC audio samples, \sys is still able to recover the correct voiceprint. In the semi-anchored and the anchored cases, the nil audio $\mathbf{0}$ is used as a placeholder when the evidence audio is unavailable.} Note that, in all three cases, the dodger to be identified or the target speaker does not have to be in the training set of \sys, i.e., \sys can handle unseen dodger and unseen target speaker.  

\section{\sys: Construction Details}
\subsection{Model Architecture Overview}
As shown in Figure~\ref{fig:model}, the voiceprint restoration model $\mathcal{E}$ consists of four blocks. 

\begin{itemize}
\setlength{\itemsep}{0pt}
\item \emph{Feature extraction \ding{192}}. The feature extraction block extracts a crude voiceprint representation from the input VC audio. 
\item \emph{Differential rectification \ding{193}}. The differential rectification block cleanses the influence of the target speaker's voiceprint by rectifying the direction of the voiceprint to be orthogonal to that of the target speaker. 
\item \emph{Dimension normalization \ding{194}}. The dimension normalization block normalizes the varied-length input into a fixed-length output. 
\item \emph{Voiceprint enhancement \ding{195}}. The voiceprint enhancement block further reduces the distance between the generated voiceprint and the ground-truth voiceprint and increases the distance between the generated voiceprint and other voiceprints, thus enhancing the discriminative capability of speaker verification/identification systems.
\end{itemize}

Note that differential rectification can only be performed with the assistance of evidence audios. Therefore, in the non-anchored case, $\mathcal{E}$ is formed by the feature extraction, dimension normalization, and voiceprint enhancement blocks only. 

\subsection{Feature Extraction}\label{subsec:featureextraction}

We leverage the state-of-the-art time delay neural network (TDNN)-based model for feature extraction~\cite{desplanques2020ecapa}. 

As shown in Figure~\ref{fig:model}, a filter bank layer first transforms time-domain audios into {``FBank'' features, which are} Mel-frequency-domain features of dimension $f\times T$, where $f$ is the number of Mel filters used and $T$ is the number of frames produced. {FBank feature is widely used as input feature for newly speaker identification systems~\cite{snyder2018x,li2017deepspeaker,desplanques2020ecapa} because it carries more information than MFCC, and is more aligned with human auditory than LPCC~\cite{joshy2016comparison}.} Then, a TDNN layer and three squeeze-excitation SE-Res2Blocks are used to extract temporal features of audios with an expanding receptive field along the time axis~\cite{gao2021res2net,yu2016multi}. A TDNN block consists of a TDNN (Conv1D) layer, followed by a rectified linear unit (ReLU) activation layer and a batch normalization (BN) layer. An SE-Res2Block consists of a TDNN layer, a Res2-Dilated-TDNN layer, another TDNN layer, and a squeeze-excitation layer. TDNNs are used to extract time-domain features, and squeeze-excitation layers model the global channel interdependencies. The convolutional kernel size and the spacing between kernel elements in the dilated convolution are denoted as $k$ and $d$ in Figure~\ref{fig:model}. Finally, the outputs of three SE-Res2Blocks are concatenated. The feature extraction block
$\mathcal{E}_1$ converts the input audio into a feature map with $3c$ channels, where $c$ is a hyperparameter. The larger the $c$ is, the more features are extracted, but the more complex the model becomes.

In the non-anchored case, the result of the feature extraction block $\mathcal{E}_1(\mathbf{x}_{vc})$ is fed directly to the dimension normalization block. In the semi-anchored and the anchored cases, the feature extraction block outputs both $\mathbf{M}= \mathcal{E}_1(\mathbf{x}_{vc})$ and $\mathbf{N}=\mathcal{E}_1(\mathbf{x}_e)$, which are fed into the differential rectification block. Note that the dimensions of $\mathbf{M}$ and $\mathbf{N}$ are $3c\times T_M$ and $3c\times T_N$ respectively. In the semi-anchored and the anchored cases, if the first input is the raw audio of suspects, the feature extraction block will output $\mathbf{M}= \mathcal{E}_1(\mathbf{x}_0)$ and $\mathbf{N}=\mathcal{E}_1(\mathbf{0})$. 


\subsection{Differential Rectification}

The differential rectification block is the key to restoring the voiceprint of the dodger with high quality by utilizing the evidence audio of the target speaker.

As mentioned in \S\ref{sec:intro}, due to the voice conversion process, the VC audio of the dodger may be close to the raw audio of the target speaker but faraway from the raw audio of the dodger in the feature space. Therefore, following the traditional feature extraction method (as \ding{192}) may not be able to generate discriminative features that are sufficient for the speaker verification/identification model to zero in on the dodger.

To tackle this problem, we propose to distill the \textit{invariant} and intrinsic features of the dodger by removing the influence of the target speaker guided by the evidence audio. To realize this goal, we develop a novel Res-Orthogonalization Block (ROB) {as shown in Figure~\ref{fig:resorth1}}, which decomposes the feature vector of the VC audio $\mathbf{M}$ into component $\mathbf{M}_{\parallel}$ that is parallel to the feature vector of the evidence audio $\mathbf{N}$ and $\mathbf{M}_{\perp}$ that is perpendicular to $\mathbf{N}$. {We present a visualization of this process in Figure~\ref{fig:resorth2}. Our intuition is that  removing the parallel component $\mathbf{M}_{\parallel}$ from $\mathbf{M}$ may suppress the influence of the target speaker induced by VC. Note that only M2 and M3 have ROB, since M1 has not been designed to handle an evidence audio.}

We first calculate the average direction of $\mathbf{N}$ along the time axis as
\begin{equation}\label{equ:direction}
\mathbf{\bar{N}} = \frac{1}{T_N}\sum_{i=1}^{T_N} \mathbf{N}_i,
\end{equation}
where $\mathbf{N}_i$ is the $i$-th frame of $\mathbf{N}$.

Then we obtain the unit direction of $\mathbf{N}$ as
\begin{equation}\label{equ:direction2}
\begin{aligned}
\mathbf{\bar{n}} = \frac{\mathbf{\bar{N}}}{|\mathbf{\bar{N}}|+\epsilon}
\end{aligned}    
\end{equation}
where the unit vector $\mathbf{\bar{n}}$ has a length of $3c$. To avoid NaN values in practice, a small constant (e.g. $\epsilon=$1e-6) is added to the denominator.

We decompose the feature vector $\mathbf{M}$ into $\mathbf{M}_{\parallel}$ and $\mathbf{M}_{\perp}$ according to $\mathbf{\bar{N}}$ as
\begin{equation}\label{equ:decompose}
\begin{aligned}
&\mathbf{M}_{\parallel}=\mathbf{\bar{n}}(\mathbf{\bar{n}}^\mathrm{T}\mathbf{M}),\\
&\mathbf{M}_{\perp}=\mathbf{M}-\mathbf{M}_{\parallel}=\mathbf{M}-\mathbf{\bar{n}}(\mathbf{\bar{n}}^\mathrm{T}\mathbf{M})=(\mathbf{I}-\mathbf{\bar{n}}\mathbf{\bar{n}}^\mathrm{T})\mathbf{M},
\end{aligned}    
\end{equation}

\noindent where $\mathbf{I}$ is an identity matrix of size $3c\times 3c$. 

Finally, we add TDNN-processed $\mathbf{M}_{\perp}$ to the original feature vector $\mathbf{M}$ in a residual learning fashion~\cite{he2016deep} as
\begin{equation}\label{equ:residual}
\begin{aligned}
\mathcal{E}_2(\mathbf{M}, \mathbf{N})=\mathrm{TDNN}(\mathbf{M}_{\perp})+\mathbf{M},\\
\end{aligned}    
\end{equation}
\noindent where $\mathcal{E}_2(\mathbf{M}, \mathbf{N})$ has the same size as $\mathbf{M}$. We show the effectiveness of our proposed differential rectification block in \S\ref{subsec:effectiveness}.

\subsection{Dimension Normalization} 

The output of the differential rectification block $\mathcal{E}_2(\mathbf{M}, \mathbf{N})$ has a dimension of $3c\times T_M$, which depends on the length of the input VC audio $t_m$. To obtain a fixed-length voiceprint,  
the dimension normalization block $\mathcal{E}_3$ applies channel- and context-dependent statistics pooling to produce an integrated representation over all channels and frames. The output of the dimension normalization block will be treated as the reconstructed voiceprint of the dodger.


\subsection{Voiceprint Enhancement}\label{subsec:enhance}
As mentioned in \S\ref{subsec:formulation}, training with normal softmax cannot yield contrasting embeddings that help speaker verification/identification models to differentiate the dodger and other suspects. Inspired by ArcFace \cite{deng2019arcface} and ECAPA-TDNN \cite{desplanques2020ecapa}, we utilize additive angular margin (AAM) loss instead of softmax loss as the output layer to explicitly force the recovered voiceprint to have high similarity with the dodger voiceprint and low similarity with the voiceprints of other suspects.

Through our experiments, we find that the entire voiceprint restoration model $\mathcal{E}$ is hard to converge during training. To boost the training efficiency, we utilize a pretrained ECAPA-TDNN model to initialize the feature extraction block $\mathcal{E}_1$. This solution enables practical training of the model with limited computing resources. 

In addition, to increase the robustness of \sys, we introduce data augmentation techniques in the training phase, including modifying speed, adding noises, randomly dropping frames. We evaluate the robustness of \sys in \S\ref{subsec:robustness}.

\section{Evaluation}\label{sec:evaluation}

\subsection{Setup}

\subsubsection{Prototype} We have implemented a prototype of \sys on the Pytorch~\cite{DBLP:conf/nips/PaszkeGMLBCKLGA19} and SpeechBrain~\cite{speechbrain} platforms and trained the model according to Equation~(\ref{equ:obj1}) using four NVIDIA 3090 GPUs. {In the training phase, both the input VC audios and the evidence audios are re-sampled to 16 kHz and randomly trimmed to 6 seconds due to the memory limitation of GPUs}, i.e., the length of the audio samples is $t_m=t_n=16,000\times6$. For the model, we set the default configuration as $c=1,024$. We detail the parameters of the blocks of the model in Table~\ref{tab:model} in the Appendix. 


In the training phase, we initialize the feature extraction block with the parameters of a pretrained ECAPA-TDNN~\cite{speechbrain} and use an Adam~\cite{kingma2014adam} optimizer to update the parameters of the model for 4$\sim$8 epochs (until a satisfying validation loss is achieved or the loss can no longer decrease), with a learning rate of 1e-3, a weight decay ($L_2$ penalty) rate of 2e-6, and a batch size of 24.

\subsubsection{Datasets and VC methods} 

To train and evaluate \sys, we generate a large-scale voice conversion dataset with {four} voice conversion methods, consisting of audios from {9,691} speakers. We implement four non-parallel one-shot voice conversion methods, i.e.,
\begin{itemize}
\setlength{\itemsep}{0pt}
    \item \textbf{VQVC}~\cite{wu2020one} is a classic disentanglement-based voice conversion method, which is based on vector quantization to disentangle speech contents and phonetic features. We utilize an unofficial implementation available on GitHub\footnote{https://github.com/Jackson-Kang/VQVC-Pytorch. Note that there is no official implementation of VQVC.}, and train the VC model on \textit{train-clean-360} of Librispeech~\cite{Panayotov2015librispeech}.
    \item \textbf{VQVC+}~\cite{wu2020vqvcp} combines VQVC and U-Net to produce high-quality audios. We utilize an official implementation available on GitHub\footnote{https://github.com/ericwudayi/SkipVQVC}, and train the VC model on \textit{train-clean-360} of Librispeech~\cite{Panayotov2015librispeech}. 
    \item \textbf{AGAIN}~\cite{chen2021again} is an improvement of the VC method AdaIN~\cite{chou2019one} with an activation guidance method. We utilize the official implementation\footnote{https://github.com/KimythAnly/AGAIN-VC}, and train the model on \textit{train-clean-360} of Librispeech.
    \item \textbf{BNE} (\textbf{BNE-Seq2seqMoL})~\cite{liu2021any} is a state-of-the-art disentanglement-based voice conversion method that has better subjective hearing quality than previous VC methods (and also shown to be the most difficult for voiceprint restoration in our experiments). We utilize both the official implementation and the official pretrained model\footnote{https://github.com/liusongxiang/ppg-vc}.
\end{itemize}

To prepare the training datasets for \sys, we apply VQVC, VQVC+, AGAIN, and BNE on three training subsets of LibriSpeech and VoxCeleb1\&2 \cite{Nagrani17, Chung18b} to generate a substantial number of VC audios. We also apply the above four VC methods to the \textit{test-clean} subset of LibriSpeech for testing. {Note that there is no overlap between any speakers (source or target) in the training set and the test set. We present the detailed information on dataset generation in Appendix~\ref{ap:dataset}.}

To evaluate the effectiveness of \sys in restoring voiceprint from audios of languages different from the language of the training set, we apply BNE on three datasets of other languages, including German, French, and Spanish. The dataset details are summarized in Table~\ref{tab:dataset} in the Appendix.

\begin{table}[tt]\centering
\resizebox{\linewidth}{!}{
\begin{threeparttable}
\setlength{\abovecaptionskip}{5pt}%
\setlength{\belowcaptionskip}{0pt}%

\caption{Performance comparison of \sys and baseline.}

\small

\setlength{\tabcolsep}{1mm}{
\begin{tabular}{@{}p{25pt}rrrrrrr@{}}
\toprule
\textbf{Method}                     &\multicolumn{1}{r}{Metrics\tnote{\textdagger}} & {B1}      & {B2}      & {B3}          & {M1}              & {M2}                         & {M3}                  \\ \midrule
\multirow{4}{*}[-4pt]{\textbf{VQVC}}  & {EER ($\downarrow$)}                              & 31.39\%   & {38.62\%}   & {47.95\%}       & 5.57\%            & 4.40\%                       & \textbf{4.08\%}                                     \\ \cmidrule(l){2-8}
                                & {Top-1~~ACC ($\uparrow$)}                         & 6.86\%    & {2.89\% }   & {3.53\% }       & 91.35\%           & 93.11\%                      & \textbf{93.21\%}                                     \\
                                & {Top-5~~ACC ($\uparrow$)}                         & 25.16\%   & {14.10\%}   & {14.10\%}       & 97.95\%           & \textbf{98.81\%}             & {98.62\%}                                           \\
                                & {Top-10~ACC ($\uparrow$)}                         & 42.89\%   & {32.05\%}   & {26.60\%}       & 99.07\%           & \textbf{99.87\%}             & {99.55\%}                                           \\ \midrule
\multirow{4}{*}[-4pt]{\textbf{VQVC+}} & {EER ($\downarrow$)}                              & 27.66\%   & {41.47\%}   & {48.97\%}       & 3.95\%            & 3.66\%                       & \textbf{3.35\%}                                     \\ \cmidrule(l){2-8}
                                & {Top-1~~ACC ($\uparrow$)}                         & 11.03\%   & {5.45\% }   & {3.85\% }       & 95.03\%           & 94.65\%                      & \textbf{95.93\%}                                     \\
                                & {Top-5~~ACC ($\uparrow$)}                         & 33.81\%   & {19.87\%}   & {17.31\%}       & 99.23\%           & 98.88\%                      & \textbf{99.36\%}                                     \\
                                & {Top-10~ACC ($\uparrow$)}                         & 54.01\%   & {33.65\%}   & {34.30\%}       & 99.65\%           & 99.55\%                      & \textbf{99.68\%}                                     \\ \midrule
\multirow{4}{*}[-4pt]{\textbf{AGAIN}} & {EER ($\downarrow$)}                              & 22.55\%   & {40.42\%}   & {42.79\%}       & 2.57\%            & 1.94\%                       & \textbf{1.87\%}                                     \\ \cmidrule(l){2-8}
                                & {Top-1~~ACC ($\uparrow$)}                         & 21.80\%   & {7.05\% }   & {4.17\% }       & 97.66\%           & 98.59\%                      & \textbf{99.07\%}                                     \\
                                & {Top-5~~ACC ($\uparrow$)}                         & 51.03\%   & {28.85\%}   & {16.67\%}       & 99.74\%           & 99.90\%                      & \textbf{99.94\%}                                     \\
                                & {Top-10~ACC ($\uparrow$)}                         & 68.88\%   & {44.55\%}   & {31.73\%}       & 99.97\%           & 99.97\%                      & \textbf{100.0\%}                                     \\ \midrule
\multirow{4}{*}[-4pt]{\textbf{BNE}}   & {EER ($\downarrow$)}                              & 31.28\%   & {45.06\%}   & {42.12\%}       & 5.14\%            & \textbf{3.58\%}              & 3.78\%                                               \\ \cmidrule(l){2-8}
                                & {Top-1~~ACC ($\uparrow$)}                         & 0.93\%    & {2.89\% }   & {1.60\% }      & 92.66\%           & 95.10\%                      & \textbf{96.12\%}                                     \\
                                & {Top-5~~ACC ($\uparrow$)}                         & 16.64\%   & {15.71\%}   & {14.42\%}       & 99.52\%           & 99.78\%                      & \textbf{99.97\%}                                     \\
                                & {Top-10~ACC ($\uparrow$)}                         & 33.33\%   & {27.89\%}   & {26.28\%}       & 99.97\%           & 99.97\%                      & \textbf{100.0\%}                                     \\ \bottomrule
\end{tabular}}

\begin{tablenotes}[flushleft]
\footnotesize
\item[\textdagger] $\uparrow$ indicates that the value is the higher, the better, and $\downarrow$ indicates that the value is the lower, the better. The best results are highlighted in bold. 

\end{tablenotes}

\label{tab:compare}
\end{threeparttable}} 
\end{table}
\begin{figure*}[tt]
    \centering
\setlength{\abovecaptionskip}{0.1cm}
\setlength{\belowcaptionskip}{0cm}
\subfigcapskip = -0.25cm
\subfigure{
    \vspace{0cm}\hspace{0.cm}\includegraphics[height=0.55cm, trim=0 155 0 5, clip]{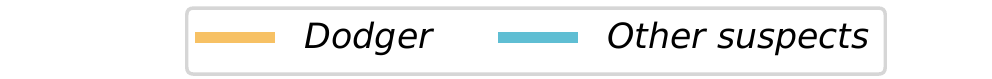}
    }
\\\vspace{-0.3cm}\setcounter{subfigure}{0}
\subfigure[VQVC]{
    \includegraphics[height=\threesize, trim=0 0 0 0, clip]{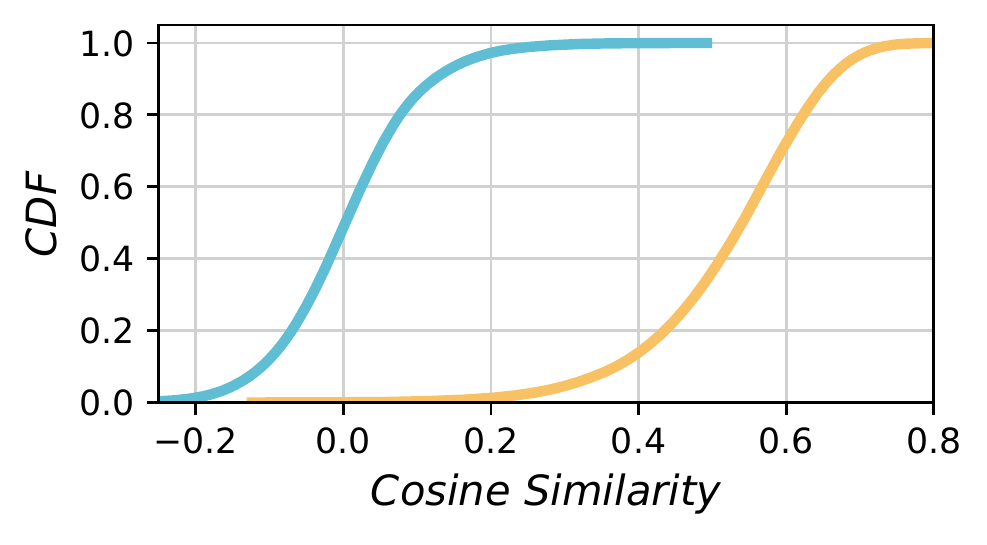}
    \label{fig:vqvc}
    }
\subfigure[VQVC+]{
    \includegraphics[height=\threesize, trim=0 0 0 0, clip]{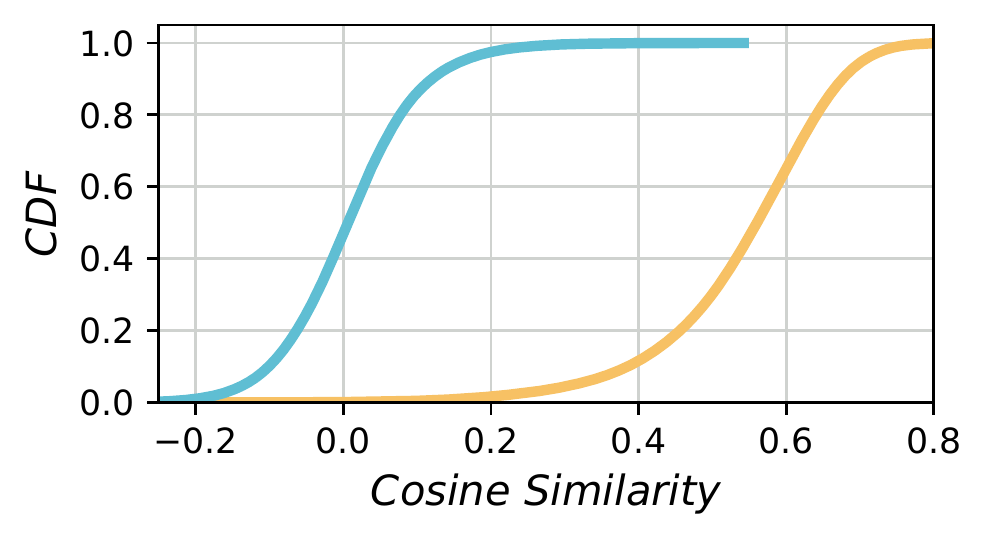}
    \label{fig:vqvcp}
    }
\subfigure[AGAIN]{
    \includegraphics[height=\threesize, trim=0 0 0 0, clip]{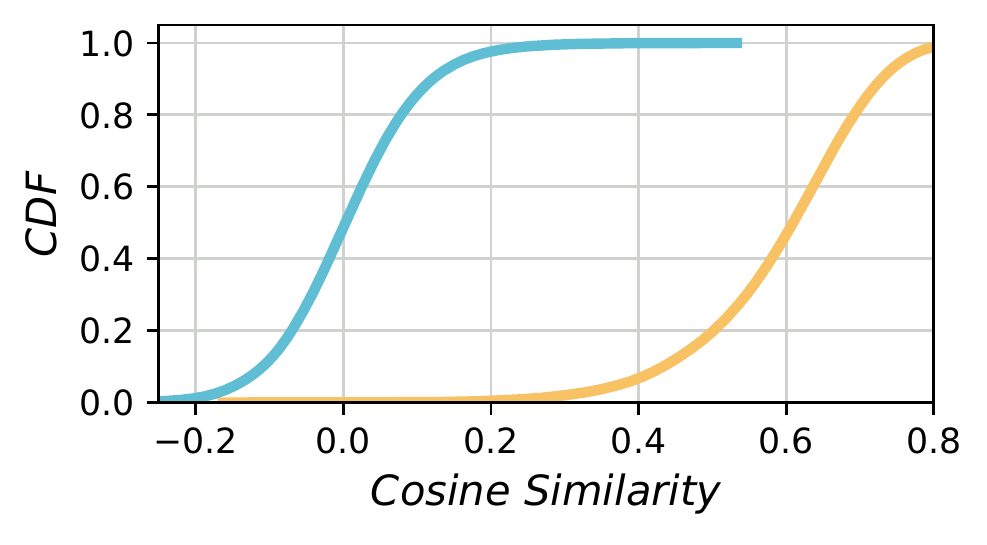}
    \label{fig:again}
    }
\subfigure[BNE]{
    \includegraphics[height=\threesize, trim=0 0 0 0, clip]{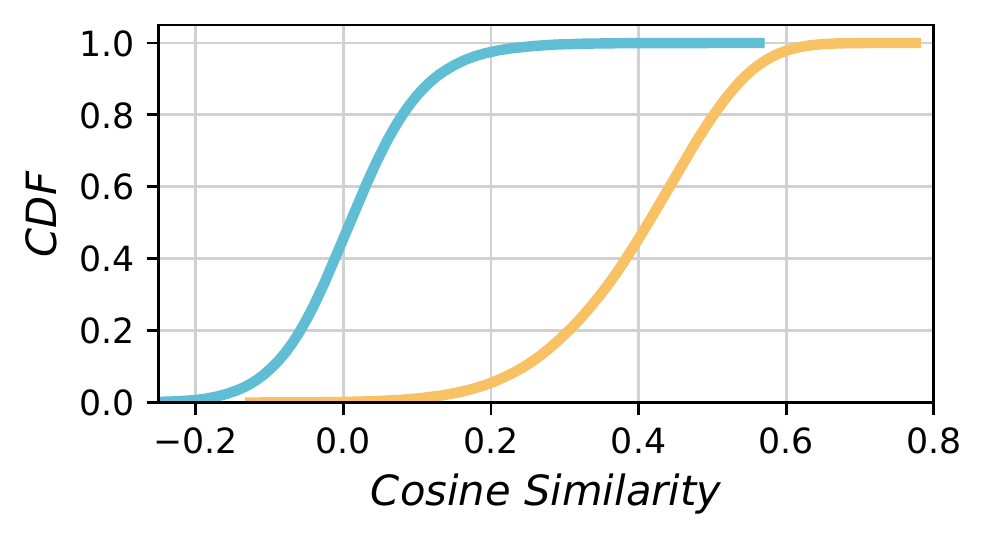}
    \label{fig:ppg}
    }

\caption{The distributions of cosine similarity scores between the restored voiceprint by \sys and the dodger or other
suspects. The voiceprints are recovered by M3.}\label{fig:dist1}
\end{figure*}
\begin{figure}[tt]
    \centering
\setlength{\abovecaptionskip}{0.cm}
\setlength{\belowcaptionskip}{0cm}
\subfigcapskip = -0.25cm
\subfigure{
    \includegraphics[width=.5\columnwidth, trim=0 0 0 0, clip]{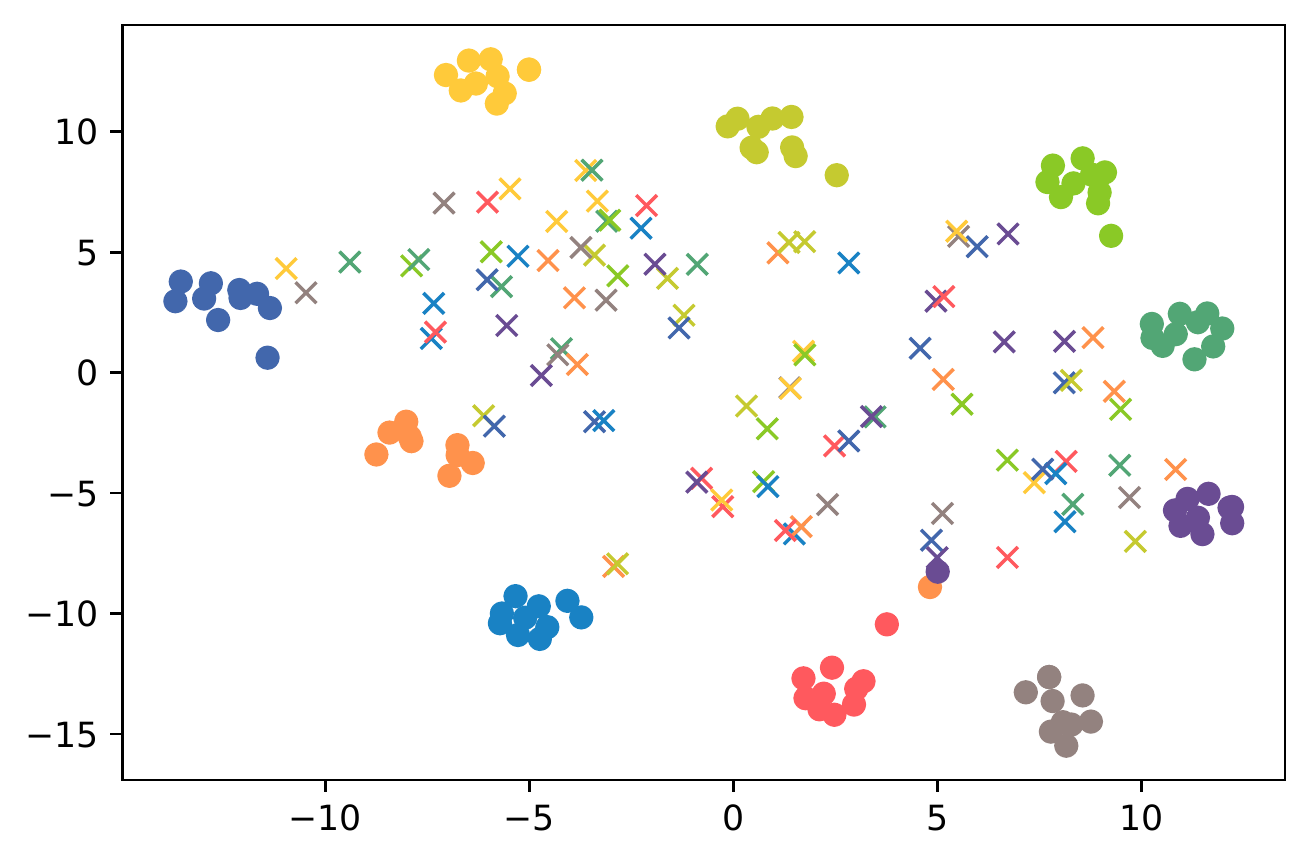}
    }
\\\vspace{-0.3cm}
\subfigure{
    \includegraphics[height=0.55cm, trim=40 260 40 6, clip]{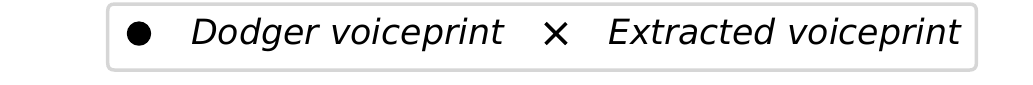}
    }
\caption{Distribution (visualized using t-SNE) of the dodger voiceprint and the extracted voiceprint by the baseline. Colors denote different dodgers. It is shown that the extracted voiceprints by the baseline are faraway from the voiceprint of the real dodger.}\label{fig:dots_baseline}
\end{figure}
\begin{figure}[tt]
    \centering
\setlength{\abovecaptionskip}{0.cm}
\setlength{\belowcaptionskip}{0cm}
\subfigcapskip = -0.25cm
\subfigure{
    \includegraphics[width=.5\columnwidth, trim=0 0 0 0, clip]{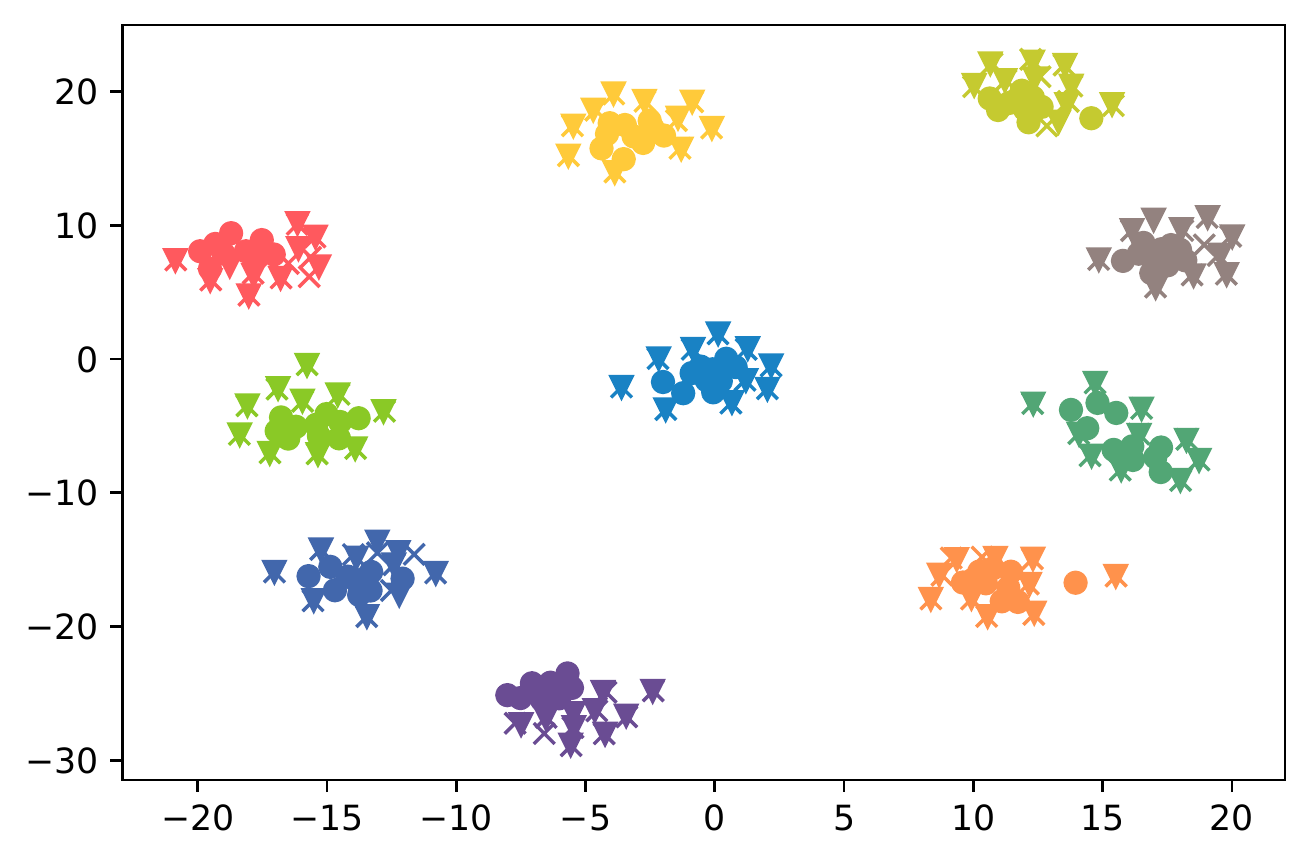}
    }
\\\vspace{-0.3cm}
\subfigure{
    \hspace{-0.3cm}\includegraphics[height=0.55cm, trim=5 270 5 6, clip]{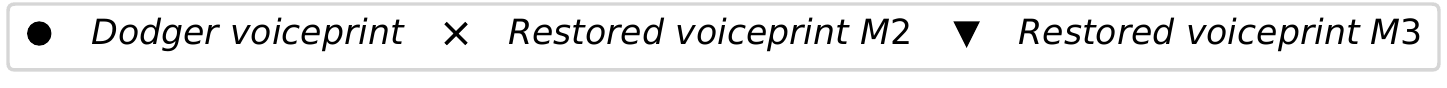}
    }
\caption{Distribution (visualized using t-SNE) of the dodger voiceprint and the restored voiceprint by \sys. Colors denote different dodgers. It is shown that the restored voiceprints by \sys are close to the voiceprint of the real dodger.}\label{fig:dots}
\end{figure}
\begin{table}[tt]\centering
\resizebox{\linewidth}{!}{
\begin{threeparttable}
\setlength{\abovecaptionskip}{5pt}%
\setlength{\belowcaptionskip}{0pt}%

\caption{Generalizability to unseen VCs.}

\small

\setlength{\tabcolsep}{3mm}{
\begin{tabular}{@{}p{25pt}rrrrr@{}}
\toprule
\multirow{2}{*}{\textbf{Method}}& \multirow{2}{*}{Metrics\tnote{\textdagger}} & \multicolumn{4}{c}{{Audios converted by}}                                                                           \\
                                &                                     & {VQVC}        & {VQVC+}           & {AGAIN}                      & {BNE~}                                         \\ \midrule
\multirow{4}{*}[-4pt]{\textbf{VQVC}}  & {EER ($\downarrow$)}                & 4.08\%            & 7.96\%            & 6.24\%                       & 36.79\%                                              \\ \cmidrule(l){2-6}
                                & {Top-1~~ACC ($\uparrow$)}           & 93.21\%           & 77.31\%           & 90.71\%                      & 4.26\%                                               \\
                                & {Top-5~~ACC ($\uparrow$)}           & 98.62\%           & 95.06\%           & 99.33\%                      & 26.35\%                                              \\
                                & {Top-10~ACC ($\uparrow$)}           & 99.55\%           & 98.08\%           & 99.78\%                      & 45.61\%                                              \\ \bottomrule
\multirow{4}{*}[-4pt]{\textbf{VQVC+}} & {EER ($\downarrow$)}                & 6.46\%            & 3.35\%            & 2.67\%                       & 35.14\%                                              \\ \cmidrule(l){2-6}
                                & {Top-1~~ACC ($\uparrow$)}           & 85.64\%           & 95.93\%           & 98.27\%                      & 3.11\%                                              \\
                                & {Top-5~~ACC ($\uparrow$)}           & 97.34\%           & 99.36\%           & 99.90\%                      & 28.27\%                                              \\
                                & {Top-10~ACC ($\uparrow$)}           & 99.10\%           & 99.68\%           & 100.0\%                      & 47.15\%                                              \\ \bottomrule
\multirow{4}{*}[-4pt]{\textbf{AGAIN}} & {EER ($\downarrow$)}                & 8.99\%            & 5.82\%            & 1.87\%                       & 31.23\%                                              \\ \cmidrule(l){2-6}
                                & {Top-1~~ACC ($\uparrow$)}           & 75.77\%           & 86.25\%           & 99.07\%                      & 2.40\%                                               \\
                                & {Top-5~~ACC ($\uparrow$)}           & 93.30\%           & 96.70\%           & 99.94\%                      & 27.44\%                                              \\
                                & {Top-10~ACC ($\uparrow$)}           & 97.82\%           & 99.33\%           & 100.0\%                      & 46.70\%                                              \\ \bottomrule
\multirow{4}{*}[-4pt]{\textbf{BNE}}   & {EER ($\downarrow$)}                & 30.35\%           & 19.84\%           & 12.11\%                      & 3.78\%                                               \\ \cmidrule(l){2-6}
                                & {Top-1~~ACC ($\uparrow$)}           & 20.77\%           & 28.21\%           & 50.42\%                      & 96.12\%                                              \\
                                & {Top-5~~ACC ($\uparrow$)}           & 52.69\%           & 64.01\%           & 83.75\%                      & 99.97\%                                              \\
                                & {Top-10~ACC ($\uparrow$)}           & 71.19\%           & 79.30\%           & 93.33\%                      & 100.0\%                                              \\ \bottomrule
\end{tabular}}

\begin{tablenotes}[flushleft]
\footnotesize
\item[\textdagger] $\uparrow$ indicates that the value is the higher, the better, and $\downarrow$ indicates that the value is the lower, the better.

\end{tablenotes}

\label{tab:unseenVC}
\end{threeparttable}}
\end{table}

\begin{table}[tt]\centering
\resizebox{\columnwidth}{!}{
\begin{threeparttable}
\setlength{\abovecaptionskip}{5pt}%
\setlength{\belowcaptionskip}{0pt}%
\caption{Impact of VC audio length.}
\small

\setlength{\tabcolsep}{3mm}{
\begin{tabular}{@{}rrrrrr@{}}
\toprule
\multicolumn{1}{r}{Metrics\tnote{\textdagger}}  & {5s}        & {10s}           & {15s}                      & {20s}                              & {25s}                 \\ \midrule
 {EER ($\downarrow$)}                           & 5.55\%      & 3.71\%          & 3.39\%                     & 3.78\%                             & 4.48\%                                              \\ \cmidrule(l){1-6}
 {Top-1~~ACC ($\uparrow$)}                      & 88.24\%     & 92.31\%         & 94.55\%                    & 96.12\%                            & 95.51\%                                              \\
 {Top-5~~ACC ($\uparrow$)}                      & 98.69\%     & 99.62\%         & 99.87\%                    & 99.97\%                            & 99.87\%                                              \\
 {Top-10~ACC ($\uparrow$)}                      & 99.62\%     & 99.87\%         & 100.0\%                    & 100.0\%                            & 100.0\%                                              \\ \bottomrule
\end{tabular}}

\begin{tablenotes}[flushleft]
\footnotesize
\item[\textdagger] $\uparrow$ indicates that the value is the higher, the better, and $\downarrow$ indicates that the value is the lower, the better.
\end{tablenotes}

\label{tab:duration}
\end{threeparttable}}
\end{table}

\begin{table}[tt]\centering
\resizebox{\linewidth}{!}{
\begin{threeparttable}
\setlength{\abovecaptionskip}{5pt}%
\setlength{\belowcaptionskip}{0pt}%
\caption{Impact of evidence audio length.}
\small

\setlength{\tabcolsep}{3mm}{
\begin{tabular}{@{}rrrrrr@{}}
\toprule
\multicolumn{1}{r}{Metrics\tnote{\textdagger}}  & {5s}        & {10s}           & {15s}                      & {20s}                              & {25s}                 \\ \midrule
 {EER ($\downarrow$)}                           & 3.70\%      & 3.67\%          & 3.68\%                   & 3.78\%                            & 3.68\%                                             \\ \cmidrule(l){1-6}
 {Top-1~~ACC ($\uparrow$)}                      & 95.61\%     & 95.67\%         & 95.80\%                  & 96.12\%                           & 95.83\%                                              \\
 {Top-5~~ACC ($\uparrow$)}                      & 99.68\%     & 99.81\%         & 99.71\%                  & 99.97\%                           & 99.84\%                                              \\
 {Top-10~ACC ($\uparrow$)}                      & 99.97\%     & 99.94\%         & 99.97\%                  & 100.0\%                           & 99.97\%                                              \\ \bottomrule
\end{tabular}}

\begin{tablenotes}[flushleft]
\footnotesize
\item[\textdagger] $\uparrow$ indicates that the value is the higher, the better, and $\downarrow$ indicates that the value is the lower, the better.
\end{tablenotes}

\label{tab:targetduration}
\end{threeparttable}}
\end{table}

\begin{table}[tt]\centering
\resizebox{\linewidth}{!}{
\begin{threeparttable}
\setlength{\abovecaptionskip}{5pt}%
\setlength{\belowcaptionskip}{0pt}%

\caption{Over-the-telephony robustness.}

\small

\setlength{\tabcolsep}{1mm}{
\begin{tabular}{@{}p{25pt}rrrrrrr@{}}
\toprule
\textbf{Method}                 &\multicolumn{1}{r}{Metrics\tnote{\textdagger}}           & {$\mu$-law}        & {A-law}           & {GSM\tnote{\textdaggerdbl}}                      & {AMR\tnote{\textdaggerdbl}}          & {8kHz\tnote{\textdaggerdbl}}                      & {4kHz\tnote{\textdaggerdbl}}        \\ \midrule
\multirow{4}{*}[-4pt]{\textbf{VQVC}}  & {EER ($\downarrow$)}                 & 5.47\%            & 5.39\%             & 7.12\%                      & 9.54\%                & {5.11\% }                     & {15.78\%}                       \\ \cmidrule(l){2-8}
                                & {Top-1~~ACC ($\uparrow$)}            & 91.60\%           & 91.76\%            & 80.32\%                     & 69.90\%               & {92.89\%}                     & {49.23\%}                       \\
                                & {Top-5~~ACC ($\uparrow$)}            & 99.58\%           & 99.68\%            & 98.49\%                     & 92.76\%               & {99.58\%}                     & {82.92\%}                       \\
                                & {Top-10~ACC ($\uparrow$)}            & 99.94\%           & 99.94\%            & 99.97\%                     & 97.47\%               & {99.84\%}                     & {92.37\%}                       \\ \bottomrule
\multirow{4}{*}[-4pt]{\textbf{VQVC+}} & {EER ($\downarrow$)}                 & 4.60\%            & 4.42\%             & 5.23\%                      & 7.68\%                & {4.56\% }                     & {22.62\%}                       \\ \cmidrule(l){2-8}
                                & {Top-1~~ACC ($\uparrow$)}            & 91.03\%           & 92.56\%            & 81.28\%                     & 77.95\%               & {91.12\%}                     & {44.90\%}                       \\
                                & {Top-5~~ACC ($\uparrow$)}            & 99.42\%           & 99.33\%            & 98.88\%                     & 96.54\%               & {99.71\%}                     & {78.91\%}                       \\
                                & {Top-10~ACC ($\uparrow$)}            & 99.94\%           & 99.87\%            & 99.78\%                     & 99.07\%               & {99.94\%}                     & {90.10\%}                       \\ \bottomrule
\multirow{4}{*}[-4pt]{\textbf{AGAIN}} & {EER ($\downarrow$)}                 & 2.08\%            & 2.15\%             & 2.63\%                      & 3.30\%                & {2.11\% }                     & {5.96\% }                       \\ \cmidrule(l){2-8}
                                & {Top-1~~ACC ($\uparrow$)}            & 98.33\%           & 97.72\%            & 96.41\%                     & 95.61\%               & {97.69\%}                     & {90.32\%}                       \\
                                & {Top-5~~ACC ($\uparrow$)}            & 100.0\%           & 99.97\%            & 99.78\%                     & 99.65\%               & {99.87\%}                     & {99.90\%}                       \\
                                & {Top-10~ACC ($\uparrow$)}            & 100.0\%           & 100.0\%            & 100.0\%                     & 100.0\%               & {100.0\%}                     & {100.0\%}                       \\ \bottomrule
\multirow{4}{*}[-4pt]{\textbf{BNE}}   & {EER ($\downarrow$)}                 & 4.27\%            & 4.16\%             & 4.87\%                      & 6.87\%                & {4.03\% }                     & {6.47\% }                       \\ \cmidrule(l){2-8}
                                & {Top-1~~ACC ($\uparrow$)}            & 92.21\%           & 94.46\%            & 90.61\%                     & 79.07\%               & {92.89\%}                     & {76.67\%}                       \\
                                & {Top-5~~ACC ($\uparrow$)}            & 99.52\%           & 99.58\%            & 99.04\%                     & 95.90\%               & {99.55\%}                     & {95.13\%}                       \\
                                & {Top-10~ACC ($\uparrow$)}            & 99.97\%           & 99.94\%            & 99.81\%                     & 98.69\%               & {100.0\%}                     & {98.08\%}                       \\ \bottomrule
\end{tabular}}             

\begin{tablenotes}[flushleft]
\footnotesize
\item[\textdaggerdbl] Short for GSM-FR, AMR-NB, 8kHz subsampling, and 4kHz subsampling.
\item[\textdagger] $\uparrow$ indicates that the value is the higher, the better, and $\downarrow$ indicates that the value is the lower, the better. 

\end{tablenotes}

\label{tab:robust}
\end{threeparttable}} 
\end{table}





\begin{table}[tt]\centering
\resizebox{\linewidth}{!}{
\begin{threeparttable}
\setlength{\abovecaptionskip}{5pt}%
\setlength{\belowcaptionskip}{0pt}%

\caption{Performance against the adaptive adversary.}

\small

\setlength{\tabcolsep}{4.5mm}{
\begin{tabular}{@{}rrrrr@{}}
\toprule
\textbf{Method}                                 & \multicolumn{4}{c}{\textbf{VC-specific M3}}                     \\
\textbf{Dataset}                                & {VQVC\tnote{$^2$}}        & {VQVC+\tnote{$^2$}}        & {AGAIN\tnote{$^2$}}         & {BNE\tnote{$^2$}~} \\ \midrule
 {EER ($\downarrow$)}                           & 3.49\%        & 3.55\%         & 2.27\%           & 7.73\%                     \\ \cmidrule(l){1-5}
 {Top-1~~ACC ($\uparrow$)}                      & 96.60\%       & 95.64\%        & 99.39\%          & 70.32\%                    \\
 {Top-5~~ACC ($\uparrow$)}                      & 99.65\%       & 99.78\%        & 100.00\%         & 91.67\%                    \\
 {Top-10~ACC ($\uparrow$)}                      & 99.90\%       & 99.97\%        & 100.00\%         & 96.31\%                    \\ \midrule
\midrule
\textbf{Method}                                 & \multicolumn{4}{c}{\textbf{Ensembled M3}}                                            \\
\textbf{Dataset}                                & {VQVC\tnote{$^2$}}        & {VQVC+\tnote{$^2$}}        & {AGAIN\tnote{$^2$}}         & {BNE\tnote{$^2$}~} \\ \midrule
 {EER ($\downarrow$)}                            & 6.34\%      & 4.95\%         & 2.99\%    & 9.84\%                             \\ \cmidrule(l){1-5}
 {Top-1~~ACC ($\uparrow$)}                       & 92.60\%     & 95.99\%        & 99.42\%   & 61.22\%                             \\
 {Top-5~~ACC ($\uparrow$)}                       & 98.81\%     & 99.55\%        & 100.0\%   & 88.88\%                             \\
 {Top-10~ACC ($\uparrow$)}                       & 99.58\%     & 99.94\%        & 100.0\%   & 96.25\%                             \\ \bottomrule
\end{tabular}}

\begin{tablenotes}[flushleft]
\footnotesize
\item[2] denotes an adaptive adversary who feeds a raw audio to a specific VC twice with two different target speakers. 

\end{tablenotes}

\label{tab:adaptive}
\end{threeparttable}} 
\end{table}

\subsubsection{Evaluation Metrics} 

Two metrics are used to evaluate the extracted voiceprint of \sys under the speaker verification model and the speaker identification model, respectively.
\begin{itemize}
\setlength{\itemsep}{0pt}
    \item \textbf{Equal Error Rate (EER)} is the rate at which false positive rate (FPR) equals false negative rate (FNR), which is a metric for the speaker verification task. A lower EER denotes that the speaker verification can verify the identity of the dodger with higher precision, i.e., a better restoration performance of \sys.
    
    \item \textbf{Top-$k$ Accuracy (Top-$k$ ACC)} is the rate at which the correct label is among the top $k$ labels predicted (ranked by similarity scores) by the speaker identification model. A higher Top-$k$ ACC means that the restored voiceprint can better narrow the scope of the dodger.
    \end{itemize}

\subsubsection{Baseline} 

We utilize the state-of-the-art voiceprint extractor, ECAPA-TDNN, trained on VoxCeleb1\&2 as our baseline B1. {We also implement two existing works~\cite{DBLP:journals/tifs/ZhengLSZZ21, DBLP:journals/dsp/WangWH15} on voice recovery, referred to as B2 and B3. We will show the baselines cannot recover the voiceprint of the source speaker of the audios processed by voice conversion.}

\subsection{Overall Effectiveness}\label{subsec:effectiveness}

In this part, we evaluate the effectiveness of \sys in restoring the voiceprint of the dodger. As shown in Table~\ref{tab:dataset}, we generate four large-scale training sets, i.e., \textit{VQ-Train}, \textit{V+-Train}, \textit{AG-Train}, \textit{BN-Train} with VQVC, VQVC+, AGAIN, and BNE respectively. {As stated in \S\ref{subsec:scenario}, we have three materializations of \sys for the non-anchored (M1), semi-anchored (M2), and anchored (M3) cases.} We train the three materializations M1$\sim$M3 on each of these four training sets respectively. Then we test our models and the baseline on the corresponding test sets, i.e., \textit{VQ-Test}, \textit{V+-Test}, \textit{AG-Test}, \textit{BN-Test}. All test audios and evidence audios are trimmed or padded to 20 seconds unless state otherwise. {It takes less than 0.1 seconds for \sys to recover one voiceprint.} Note that the speakers in the test set and the training set have no overlap.

As shown in Table~\ref{tab:compare} (the best results highlighted), \sys achieves an average EER of 4.31\% (M1), 3.40\% (M2) and 3.27\% (M3) on the four VCs. Note that the average EER of the baseline B1 is as high as 28.22\%. Compared with B1, the results show that \sys can effectively recover voiceprints of the dodger from VC audios, while the plain voiceprint extractor cannot. The reason is that plain voiceprint extractors tend to extract features that are related to the target speakers introduced by VC techniques. In comparison, \sys suppresses the influence of the target speaker. {As shown in Table~\ref{tab:compare}, B2 and B3 fail to recover voiceprint from VC audios, since they are restricted to traditional frequency-domain voice transformation but not learning-based voice transformation. } Note that, in Table~\ref{tab:compare}, we use 5-second test audios and evidence audios for VQVC, VQVC+ and AGAIN to show the differences among three materializations of \sys. Further evaluation of audio length is presented in \S\ref{subsec:length}.

{We visualize the cumulative distribution functions (CDF) of the similarity scores between the restored voiceprint and that of the real dodger and other suspects in Figure~\ref{fig:dist1} (\sys) and Figure~\ref{fig:distbaseline} in the Appendix (baseline B1). The scores are calculated from 31,200 pairs of audios from the same speaker (the VC audio whose source speaker is the same as the non-VC audio) and 31,200 pairs of audios from different speakers (the VC audio whose source speaker is different from the non-VC audio).} It shows that for \sys, the similarity scores between the restored voiceprint and that of the dodger are much higher than those between the restored voiceprint and other suspects, making the two CDFs easily distinguishable with a preset threshold. However, for the baseline voiceprint extractor, the distributions of the real dodger and other suspects are too close to be separated, justifying the effectiveness of \sys. {From Figure~\ref{fig:dist1}, we can see that the distribution gaps of VQVC, VQVC+, and AGAIN are more distinct than those of BNE, meaning that the VC audios produced by BNE are harder to recover. The potential reason is that BNE has the best voice conversion quality, which indicates that BNE removes the phonetic features of the dodger \textit{more thoroughly}, thus the distributions of BNE are less distinguishable than the other three VCs. It indicates that \sys may be leveraged as a way to evaluate the performance of voice conversion techniques.}

As for the speaker identification task, as shown in Table~\ref{tab:compare}, we can see that \sys achieves an average Top-1 ACC of 94.18\% (M1), 95.36\% (M2), and 96.08\% (M3) on the four VCs. \sys achieves an average Top-5 ACC of 99.11\% (M1), 99.34\% (M2), and 99.47\% (M3) on the four VCs respectively. We also visualize the original voiceprints of 10 speakers (i.e., the dodgers) and the corresponding extracted voiceprints by the baseline model in Figure~\ref{fig:dots_baseline}, and the recovered voiceprints by M2 and M3 in Figure~\ref{fig:dots} using t-SNE~\cite{Maaten2008visualizing}. We can see that the original voiceprint and the recovered voiceprint of the same speaker are tightly clustered. The results imply that the detector can utilize \sys to narrow down the suspects and even pinpoint the dodger directly with a high probability. {Note that we do not present the results of M1 in Figure~\ref{fig:dots} because M1 and M2 (or M3) are different models and they map the same speaker into different voiceprints.}

The results in Table~\ref{tab:compare} also verify the effectiveness of our proposed differential rectification block, since M2 and M3 outperform M1 in most cases. In addition, the performances of M3 are better than M2 in most cases, which implies that the model indeed leverages the extra information of the target speaker to improve the voiceprint restoration ability.

\subsection{{Generalizability to Unseen VCs}}
{In this part, we evaluate the generalizability of \sys to VC methods that are unseen in the training phase. We utilize M3 trained on one VC to identify the source speaker of VC audios generated by the other three VCs respectively. We present the results in Table~\ref{tab:unseenVC}. The models trained on VQVC, VQVC+ and AGAIN can generalize well to each other, e.g., the model trained on VQVC achieves an EER of 7.96\% (resp. 6.24\%) and a Top-1 ACC of 77.31\% (resp. 90.71\%) on VC audios generated by VQVC+ (resp. AGAIN). A possible reason is that these three VC methods share a similar architecture for disentanglement. Due to significant differences in architecture, these three models can not generalize well to BNE, vice-versa. But the model trained on BNE can still narrow down the dodger on VC audios converted by VQVC, VQVC+ and AGAIN with a Top-10 ACC of 71.19\%, 79.30\%, and 93.33\% respectively.
}

\subsection{Intra- \& Inter-gender Performance}

In this part, we evaluate the voiceprint restoration performance of M3 on different types of dodger-target pairs, i.e., male-to-male (M$\rightarrow$M), female-to-female (F$\rightarrow$F), female-to-male (F$\rightarrow$M), and male-to-female (M$\rightarrow$F). The four test sets are split from \textit{BN-Test}. As shown in Table~\ref{tab:gender}, we can see that the performances of intra- and inter-gender cases are similar, with an average EER of 2.69\% (M$\rightarrow$M), 3.72\% (F$\rightarrow$F), 3.90\% (F$\rightarrow$M), and 3.73\% (M$\rightarrow$F) respectively. The same is true for the speaker identification performance. Note that we still perform speaker identification among all 40 speakers. The results imply that \sys can handle both intra- and inter-gender conversion even if the intra- and inter-gender conversion may induce different intensities of distortions on the dodger voiceprint.

\subsection{Audio Length}\label{subsec:length}
In this part, we first examine how long the VC audio needs to be to recover the dodger voiceprint. We trim BNE-processed VC audios to 5s, 10s, 15s, 20s, and 25s. As shown in Table~\ref{tab:duration}, we find that a 10-second VC audio is sufficient for voiceprint restoration, achieving an EER of 3.71\% and a Top-1 ACC of 92.31\%. This requirement is practical in the real world since a phone scam can hardly be completed in less than 10 seconds. In addition, the results imply that a longer VC audio is beneficial to voiceprint restoration. {Note that in Table~\ref{tab:duration}, the performance decreases at 20s and 25s because some of the VC audios in the dataset have a shorter length and are padded to 20s and 25s with zeros, which affects the restoration.
}

Moreover, we examine how long the evidence audio needs to be to help recover the dodger voiceprint. We trim evidence audios to 5s, 10s, 15s, 20s, 25s and see the performances of voiceprint restoration. As shown in Table~\ref{tab:targetduration}, we find that a 5-second evidence audio is sufficient for improving the voiceprint restoration performance, yielding a Top-1 ACC of 95.61\%. In comparison, with no evidence audio, Table~\ref{tab:compare} shows that \sys achieves a Top-1 ACC of 95.10\%. 
This implies that the evidence audio can help recover the dodger voiceprint.

\subsection{Unseen Language}

In this part, we evaluate whether \sys can generalize to VC audios in other languages that are unseen in the training phase. We convert German, French, and Spanish test sets of multilingual LibriSpeech using BNE and evaluate the performance of M3. As shown in Table~\ref{tab:language} in the Appendix, \sys achieves an EER of 6.08\%, 5.78\%, 3.68\% and a Top-1 ACC of 80.40\%, 64.71\%, 92.11\% on German, French, and Spanish test sets respectively. The results show that \sys trained on English training sets can generalize to German, French, and Spanish audios, especially Spanish audios, on which \sys achieves comparable performances as on English audios. This language generalization capability is useful in multilingual scenarios. For example, the VC audio targets a Spanish-speaking victim living in English-speaking countries. In this case, the local detector with a \sys model trained with English corpus can also recover the voiceprint of the dodger in the Spanish-speaking VC audio.

\subsection{Performance over Telephony}\label{subsec:robustness}

{In this part, we evaluate whether \sys can recover the voiceprint of the dodger if the VC audio is encoded by telephony codecs, especially those used in a public switched telephone network (PSTN) or a voice over Internet protocol (VoIP) network. We apply four commonly-used codecs and 8k/4kHz subsampling on the VC audios to simulate the coding and transmission telephony channel.}

\begin{itemize}
\setlength{\itemsep}{0pt}
    \item G.711 ($\mu$-law) and G.711 (A-law) \cite{G711} are companding algorithms primarily used in 8-bit PCM digital telecommunication systems to optimize the dynamic range of an analog signal for digitization. $\mu$-law is commonly used in North America and Japan, while A-law is used in Europe.
    \item GSM Full Rate (GSM-FR) \cite{ETSI} is the first digital speech coding standard in the GSM digital mobile phone system. The speech encoder accepts 13-bit linear PCM at an 8 kHz sampling rate.
    \item Adaptive Multi-Rate (AMR-NB or GSM-AMR) \cite{AMR} consists of a multi-rate narrowband speech codec that encodes narrowband (200–3400 Hz) signals. AMR-NB is widely used in GSM and UMTS.
\end{itemize}

{Audios converted by VQVC, VQVC+, AGAIN, and BNE are encoded using these four codecs or subsampled to 8k/4kHz, and we evaluate whether the corresponding M3 models can still perform voiceprint restoration. As shown in Table~\ref{tab:robust}, \sys achieves an average EER of 4.11\%, 4.03\%, 4.96\%, 6.85\%, 3.95\%, and 12.71\% on recovering audios distorted by $\mu$-law, A-law, GSM-FR, AMR-NB and 8k/4kHz subsampling. \sys almost maintains its original effectiveness against $\mu$-law, A-law and 8kHz subsampling as compared to recovering non-codec/non-subsampled audios. \sys still maintains acceptable performances under GSM-FR (3.27\%$\rightarrow$4.96\%) and AMR-NB (3.27\%$\rightarrow$6.85\%), which might result from the data augmentation techniques we use in the training phase. The performance degradation on VQVC and VQVC+ audios subsampled to 4kHz is due to the significant distortion brought by subsampling. Even so, \sys can still narrow down the dodger with a Top-5 ACC of 82.92\%, 78.91\%, 99.90\% and 95.13\% from subsampled VQVC, VQVC+, AGAIN and BNE audios respectively.}

\subsection{Model Ensembling}
In this part, we evaluate whether one model can learn to restore voiceprints processed by different VCs. We sample one-fifth of the audios from the four VC training sets and the raw training sets respectively and make a multi-VC training dataset. Then we train \sys on the multi-VC dataset. As shown in Table~\ref{tab:mvc} in the Appendix, \sys achieves an EER of 2.94\%, 2.22\%, 1.65\%, 4.66\% and a Top-1 ACC of 99.84\%, 99.94\%, 99.97\% and 90.22\% on the test sets processed by VQVC, VQVC+, AGAIN, and BNE respectively. We visualize the distributions of the similarity scores between the restored voiceprint and that of the real dodger and other suspects in Figure~\ref{fig:dist2} in the Appendix. We can see that the distributions are similar to those shown in Figure~\ref{fig:dist1}, presenting an obvious gap between the real dodger and other suspects. It shows that the ensemble model can learn to recover voiceprint from four different VCs with promising performances. The detector can train an ensemble \sys model on a multi-VC dataset generated with all popular VC techniques, thus being able to recover the voiceprint of most VC audios. We also evaluate the over-the-telephony robustness of the ensemble model, as shown in Table~\ref{tab:robust_mvc} in the Appendix. We can see that the ensemble model still maintains acceptable performances in most cases. 

\subsection{{Adaptive Adversary}}
{
In this part, we consider an adaptive adversary who feeds a raw audio to VC twice with two different target speakers, e.g., Alice first converts the raw audio to sound like Bob and then converts the converted audio to sound like Eve, both by the same VC. In such a case, Alice is the source speaker to be identified and Eve is the final target speaker of the evidence audio. As shown in Table~\ref{tab:adaptive}, \sys still achieves very low EER and very high ACC on the audios that are converted twice by VQVC, VQVC+, and AGAIN. The performance on BNE audios is not as good as others. The possible reason is that BNE's disentanglement module is better at erasing the voice characteristics of the source speaker in the process of two conversions. For BNE, \sys can still achieve an EER of 7.73\%, a Top-1 ACC of 70.32\% and a Top-5 ACC of 91.96\%.
}
\section{Related Work}{

\subsection{Voice Conversion (VC)}
VC techniques can be categorized into parallel VC and non-parallel VC according to whether they require parallel corpus for training.

\textbf{Parallel VC}.  
Parallel VC leverages parallel corpus for training the VC model. Abe et al.~\cite{abe1988voice} created code vectors based on hard clustering using vector quantization (VQ) for feature mapping. Several subsequent works~\cite{shikano1991speaker, arslan1997voice, turk2006robust} tried to reduce the quantization error of VQ by using a fuzzy VQ based on soft clustering and the output is weighed according to the source speaker features. Wu et al.~\cite{wu2013exemplar} output a linear combination of the exemplars with weights determined by the source speaker features. Stylianou et al.~\cite{stylianou1998continuous} utilized a continuous probabilistic transformation method based on Gaussian mixture models (GMMs). Toda et al.~\cite{toda2007voice} alleviated the over-smoothing effect of the GMM-based method with global variance. Nakashika et al.~\cite{nakashika2013voice} modeled the source and the target speaker with deep belief nets (DBNs) and converted the audio using a neural network (NN). Sun et al.~\cite{sun2015voice} used a bidirectional long short-term memory-based recurrent neural network (BiLSTM) for conversion.

\textbf{Non-Parallel VC}. To leverage non-parallel corpora, several attempts~\cite{zhang2008text, erro2010inca, song2013nonparallel, benisty2014nonparallel} have been made to adapt parallel VC to use non-parallel training sets, but mismatches in alignment may occur. 
To avoid explicit alignment, Hsu et al.~\cite{hsu2016voice} used a variational auto-encoder (VAE) for speech feature transformation. Later works proposed different models for transformation~\cite{nakashika2016nonparallel,  hsu2017voice, saito2018nonparallel, kaneko2018cyclegan, fang2018high, tobing2019nonparallel, kameoka2018stargan}, but fail to handle target speakers that are unseen in the training process. To tackle this shortcoming, advanced disentanglement-based methods have been proposed. Chou et al.\cite{chou2019one} disentangled speaker and content features with an average pooling layer and the instance normalization technique (IN), and synthesized them with the adaptive instance normalization technique (AdaIN). Wu et al.~\cite{wu2020one} proposed VQVC, utilizing VQ and IN for disentanglement. Later, they combined VQ with U-Net to produce high-quality audios~\cite{wu2020vqvcp}. Chen et al.~\cite{chen2021again} developed AGAIN that uses activation as an information bottleneck to further separate speaker features from content features. Liu et al.\cite{liu2021any} extracted content features with an encoder-decoder-based hybrid connectionist-temporal-classification-attention (CTC-attention) phoneme recognizer. Disentanglement-based methods have become a trend of real-time VC because non-parallel and one-shot conversion can be achieved. 

\subsection{Voice Conversion Detection}

Speech deepfake detection is one of the aims of the ASVspoof 2021 Challenge~\cite{yamagishi2021asvspoof} and is the main topic of the 2022 Audio Deep Synthesis Detection (ADD) Challenge~\cite{yi2022add}. VC is a large part of deepfake techniques. Many efforts have been devoted to VC detection, e.g., spectro-temporal graph attention network (GAT)~\cite{tak2021end}, differentiable architecture~\cite{ge2021raw}, a cascade of an embedding extractor~\cite{chen2021pindrop}, continual learning~\cite{ma2021continual}, and multi-task learning~\cite{zhang2021multi}. Several attempts have been made into dataset generation~\cite{yi2021half} and data augmentation~\cite{das2021know, chen2021ur, tomilov2021stc, chen2020generalization} for model generalization. Nonetheless, existing works only aim at detecting whether the audio is genuine or fake, while we take a step further to restore the voiceprint of the source speaker of a VC audio. 
}

\section{{Responsible Disclosure}}




{There are both legitimate and illegitimate reasons to use VCs. The rest of this paper discusses illegitimate; legitimate reasons include voice anonymization for privacy concern, voice dubbing for one who is unavailable, aids for the speech-impaired, etc.} Any system that can determine the identity of a person behind a VC has the potential to harm those who are using it for legitimate purposes and has the ability to find those who are using it for illegitimate purposes. Having weighed the benefits and harms, we are releasing \sys in a limited way, i.e., we provide the code of \sys upon request to, e.g., law enforcement or professors at other institutions who are doing related research. This helps with providing the code to those who can use it for legitimate purposes, while reducing the potential harms from releasing it publicly. In addition, we have contacted companies that provide voice conversion services, reported the issues that might be brought by \sys, and recommended that they inform users about the privacy concerns. 

\section{Discussion}

In this section, we discuss the limitations of \sys.


{\emph{Adaptive voice conversion}. Being aware of \sys, a dodger may adapt the voice conversion process to compromise the voiceprint restoration effort of \sys, e.g., converting to a target speaker twice with two different VCs. Our preliminary experiments show that the recovery results of VC combinations, e.g., AGAIN(BNE($\mathbf{x}_0$)), are not ideal but still achieve an EER of 17.9\% and a Top-1 ACC of 57.9\%.} 
{The possible reason is that a combination of VCs can be seen as a new kind of VC, to which \sys fails to generalize.  A possible solution is to build a generative adversarial network (GAN) to simulate the game between voice disguise and voiceprint recovery. We consider this an interesting future direction.
}

{\emph{Inverse voice conversion}. With the recovered voiceprint of the dodger, it is possible to reconstruct the raw audio before voice conversion. Therefore, we may perform an end-to-end recovery with the converted audio as the input and the recovered raw audio as the output. Nonetheless, existing VCs require a raw audio as input and extract the speaker embeddings internally. It is difficult to align the speaker embedding recovered by \sys with the one used in VC as we only have black-box access to the VC model. In the future, we will explore this direction.}

{\textit{VC quality.} In \S\ref{subsec:effectiveness}, we have found that the VC quality impacts the performance of voiceprint recovery. A better VC (like BNE) performs a more thorough conversion to the target speaker, thus preserves less information of the source speaker. Nonetheless, there is no existing VC technique that can perfectly disentangle content and phonetic features, which leaves space for voiceprint restoration.}

\section{Conclusion}
In this paper, we present the rationale, design and evaluation of \sys, a voiceprint restoration model that attempts to retrieve the voiceprint of the source speaker from audios processed by voice conversion techniques. We have established a representation learning model that can effectively learn the voiceprint embedding of the source speaker given the input VC audios. With the target speaker evidence audio, we have designed a novel differential rectification to further refine the extracted voiceprint. Our experiments have confirmed the effectiveness of \sys in rebuilding voiceprint under various voice conversion techniques. 

\section*{Acknowledgments}
We sincerely thank our Shepherd and all the anonymous reviewers for their valuable comments. This work is supported by China NSFC Grant 61925109. Yanjiao Chen is the corresponding author.

\clearpage
\bibliographystyle{plain}
\bibliography{main-usenix}
\appendix

\section{{Dataset Generation}}\label{ap:dataset}
{As shown in Table~\ref{tab:dataset}, we have nine clean audio datasets for voice conversion dataset generation, i.e., \textit{train-clean-100}, \textit{train-clean-360}, \textit{train-other-500}, VoxCeleb1, VoxCeleb2, \textit{test-clean}, MLS German, MLS French, and MLS Spanish.}

\begin{itemize}
    \item \textit{train-clean-100} has 251 speakers. For each source speaker in \textit{train-clean-100}, the remaining 250 speakers are used as the target speaker to generate VC samples. 
    
    \item For each source speaker in \textit{train-clean-360}, \textit{train-other-500}, and VoxCeleb1, we randomly sample another 300 speakers in the same dataset as the target speaker.
    
    \item For each source speaker in VoxCeleb2, we randomly sample another 100 speaker as the target speaker. 
\end{itemize}  
{We generate one VC sample for each of the above source-target pair for training.}

\begin{itemize}
    \item \textit{test-clean} has 40 speakers. For each source speaker, the remaining 39 speakers in the test set are used to generate test samples. We generate 20 samples for each source-target pair (a total of $40\times39\times20=$31,200 samples as shown in Table~\ref{tab:dataset}). The speaker identification system is enrolled with all 40 speakers for evaluations.
    
    \item MLS German, MLS French, and MLS Spanish have 30, 18, 20 speakers respectively. For each source speaker, all remaining speakers in the dataset are used to generate test samples. We generate 20 samples for each source-target pair.
\end{itemize}

\begin{figure*}[hb]
    \centering
\setlength{\abovecaptionskip}{0.1cm}
\setlength{\belowcaptionskip}{-0.2cm}
\subfigcapskip = -0.cm
\subfigure{
    \vspace{0cm}\hspace{0.cm}\includegraphics[height=0.55cm, trim=30 505 70 5, clip]{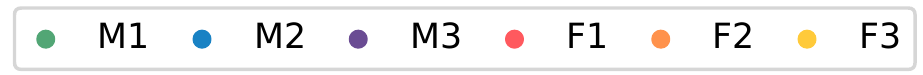}
    }
\\\vspace{-0.5cm}\setcounter{subfigure}{0}
\subfigure[VQVC]{
    \includegraphics[height=\theosize, trim=0 0 0 0, clip]{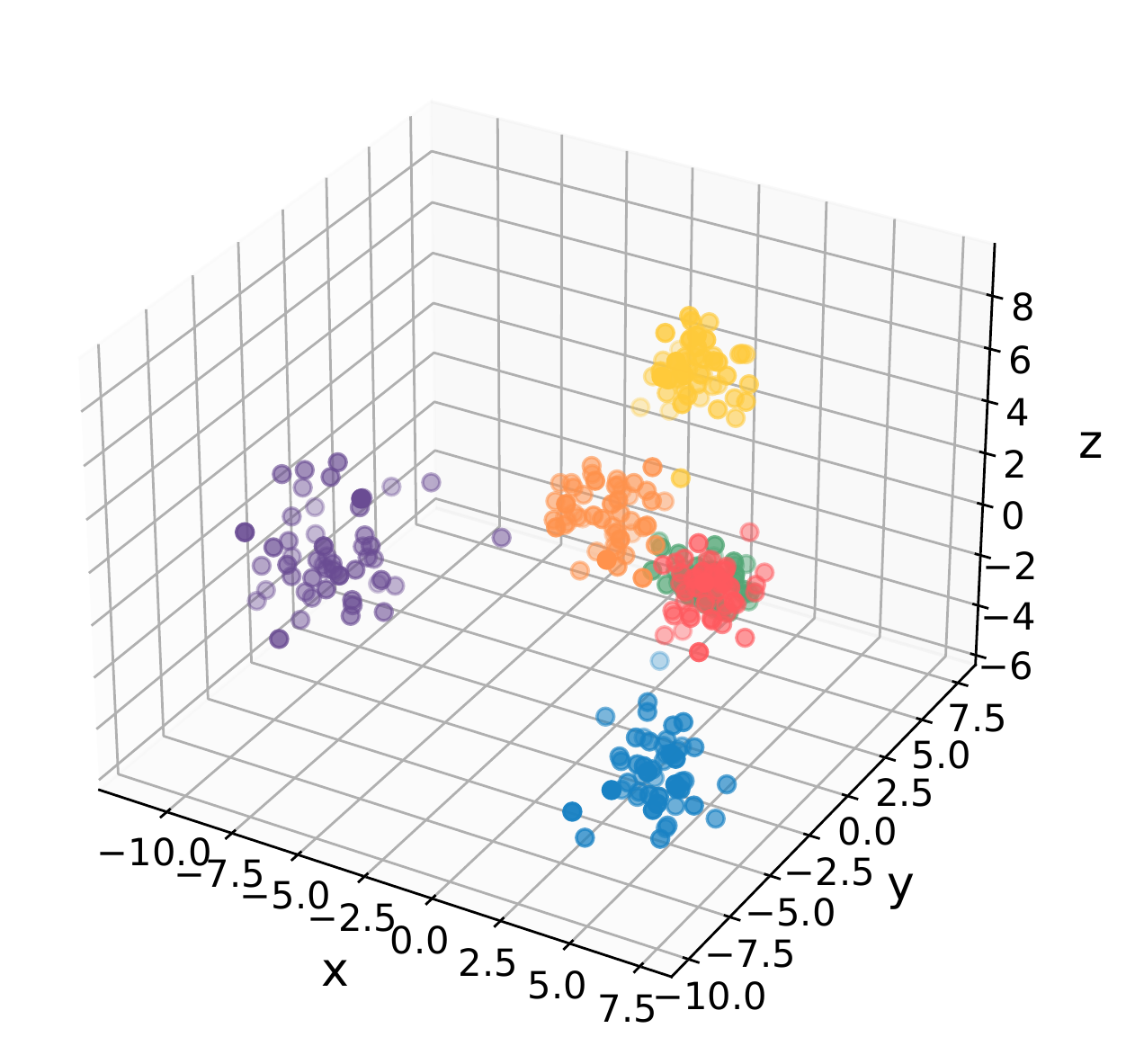}
    \label{fig:vqvc-theoretical}
    }
\subfigure[VQVC+]{
    \includegraphics[height=\theosize, trim=0 0 0 0, clip]{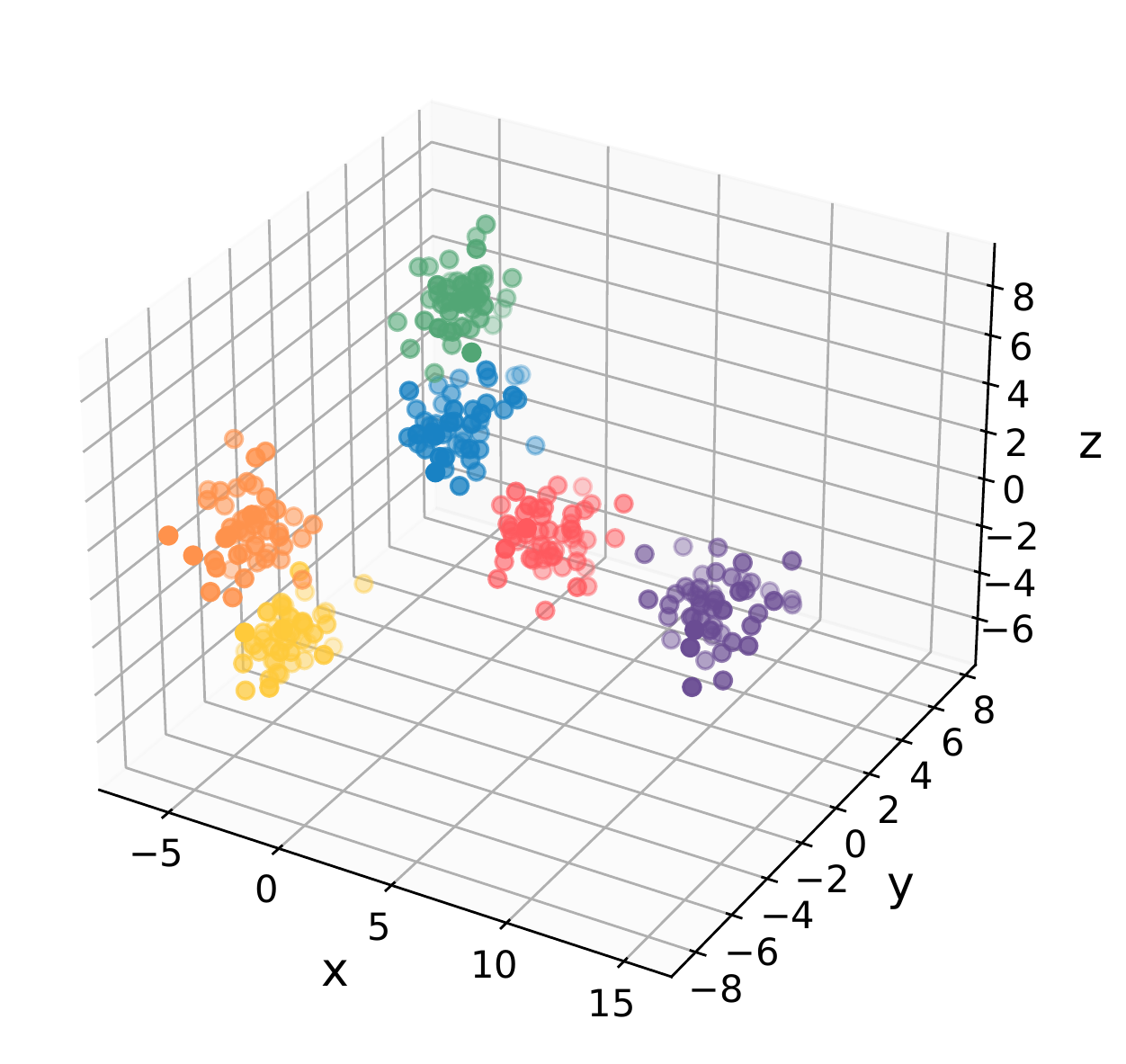}
    \label{fig:vqvcp-theoretical}
    }
\subfigure[AGAIN]{
    \includegraphics[height=\theosize, trim=0 0 0 0, clip]{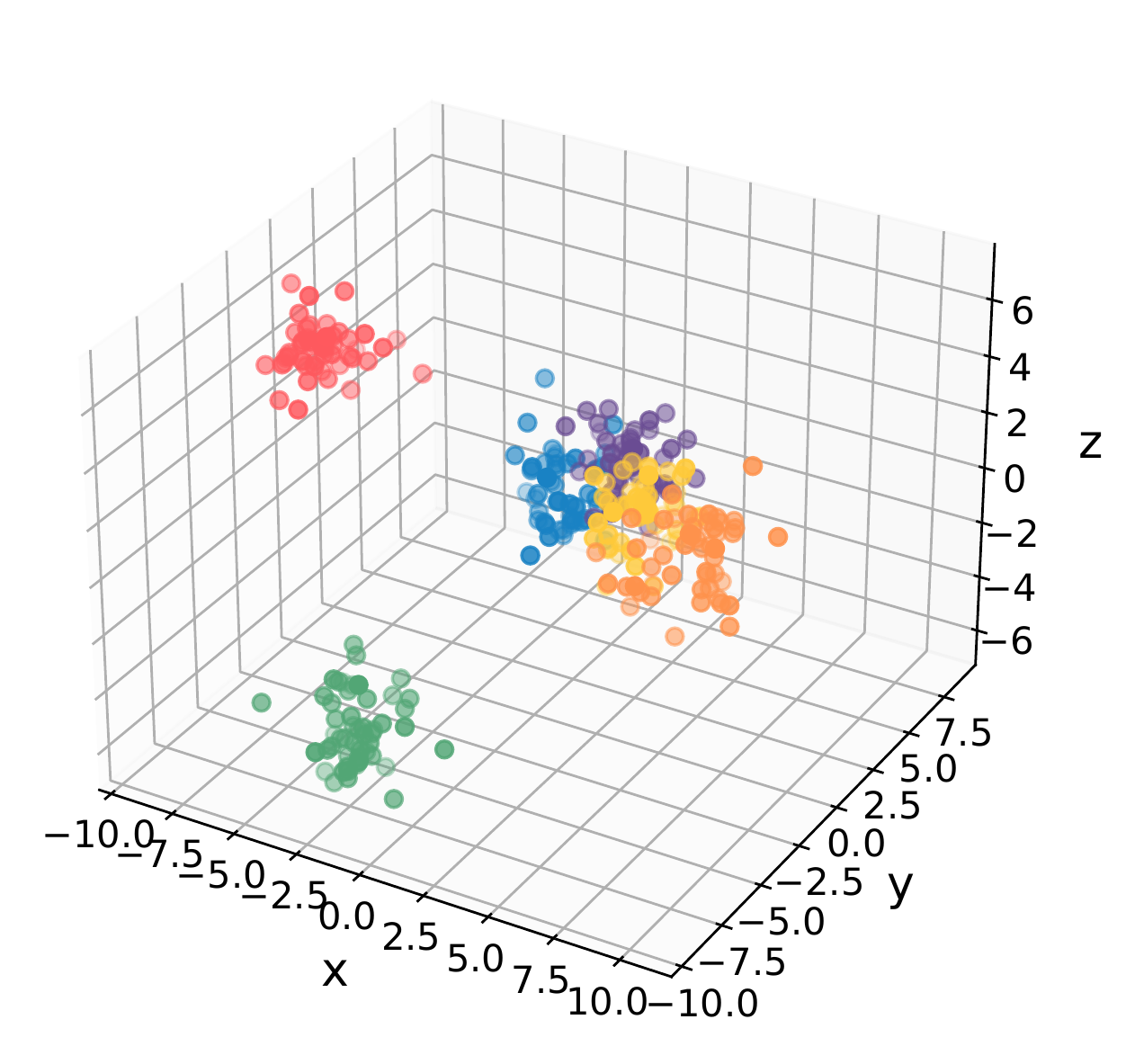}
    \label{fig:again-theoretical}
    }

\caption{Disentangled content samples of different speakers by (a) VQVC~\cite{wu2020one}, (b)  VQVC+~\cite{wu2020vqvcp} and, (c) AGAIN~\cite{chen2021again}. }\label{fig:theoretical}
\end{figure*}
\begin{figure*}[tt]
    \centering
\setlength{\abovecaptionskip}{0.1cm}
\setlength{\belowcaptionskip}{-0.2cm}
\subfigcapskip = -0.25cm
\subfigure{
    \vspace{0cm}\hspace{0.cm}\includegraphics[height=0.55cm, trim=0 155 0 5, clip]{figures/legend.pdf}
    }
\\\vspace{-0.3cm}\setcounter{subfigure}{0}
\subfigure[VQVC]{
    \includegraphics[height=\threesize, trim=0 0 0 0, clip]{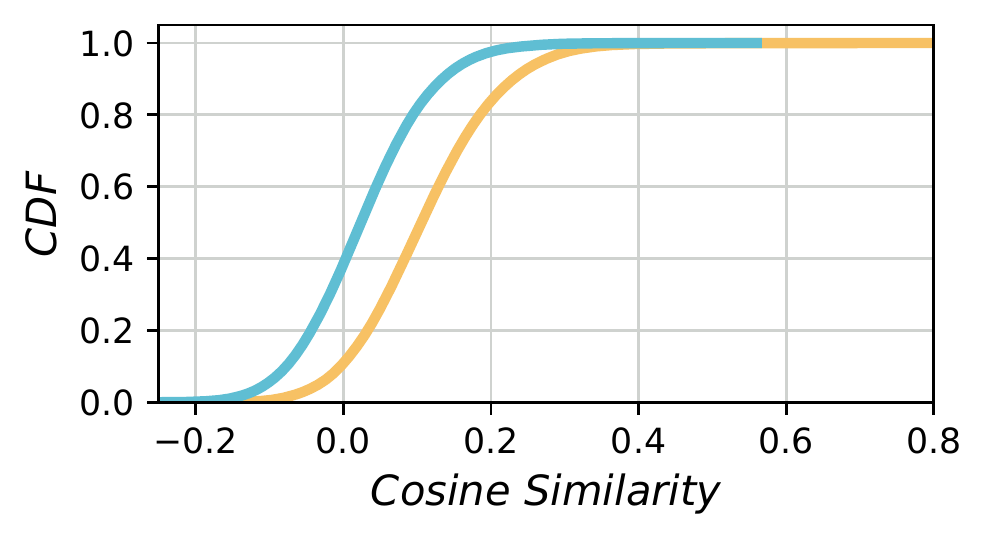}
    \label{fig:vqvc-baseline}
    }
\subfigure[VQVC+]{
    \includegraphics[height=\threesize, trim=0 0 0 0, clip]{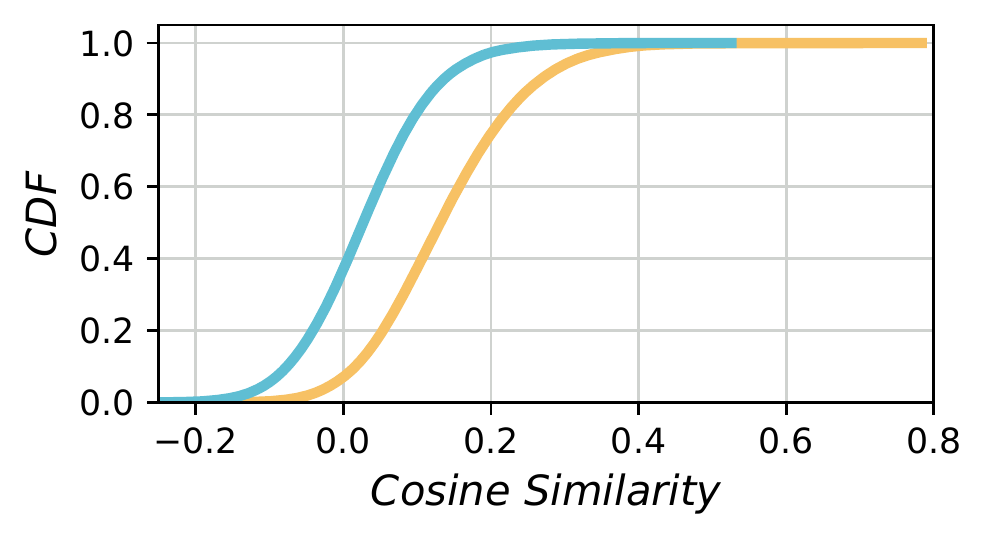}
    \label{fig:vqvcp-baseline}
    }
\subfigure[AGAIN]{
    \includegraphics[height=\threesize, trim=0 0 0 0, clip]{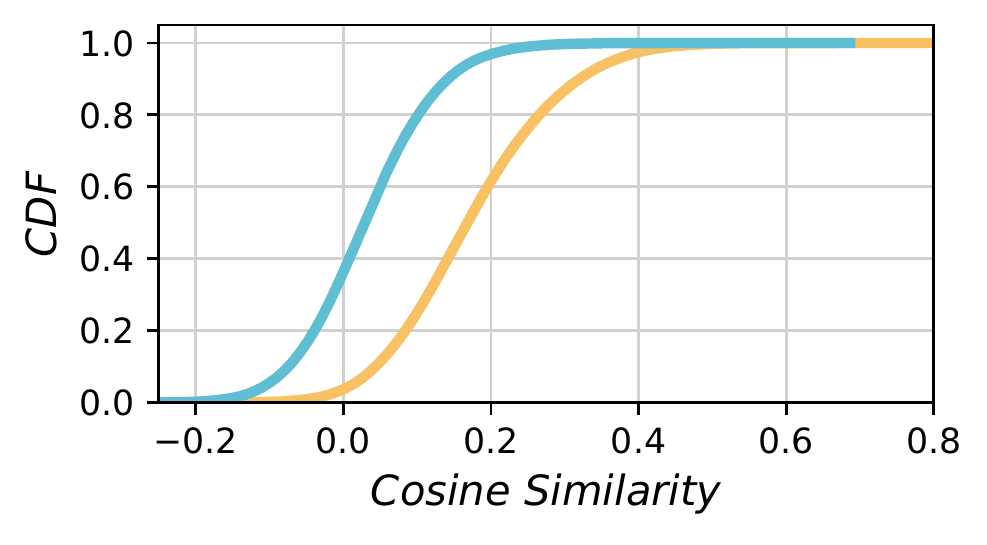}
    \label{fig:again-baseline}
    }
\subfigure[BNE]{
    \includegraphics[height=\threesize, trim=0 0 0 0, clip]{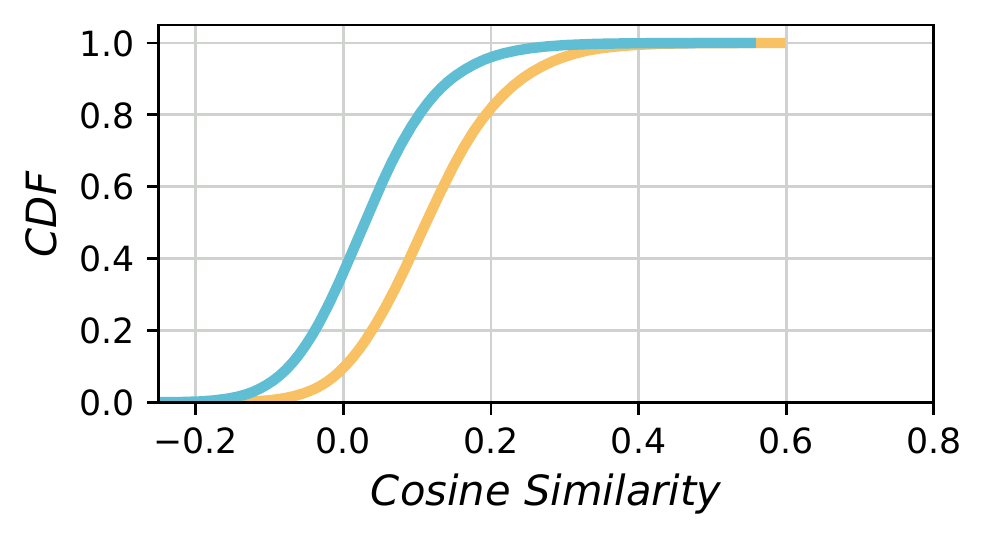}
    \label{fig:ppg-baseline}
    }

\caption{The distributions of cosine similarity scores between the extracted voiceprint by the baseline and the dodger or other suspects. }\label{fig:distbaseline}
\end{figure*}
\begin{figure*}[hb]
    \centering
\setlength{\abovecaptionskip}{0.5cm}
\setlength{\belowcaptionskip}{-0.2cm}
\subfigcapskip = -0.25cm
\subfigure{
    \vspace{0cm}\hspace{0.cm}\includegraphics[height=0.55cm, trim=0 155 0 5, clip]{figures/legend.pdf}
    }
\\\vspace{-0.3cm}\setcounter{subfigure}{0}
\subfigure[VQVC]{
    \includegraphics[height=\threesize, trim=0 0 0 0, clip]{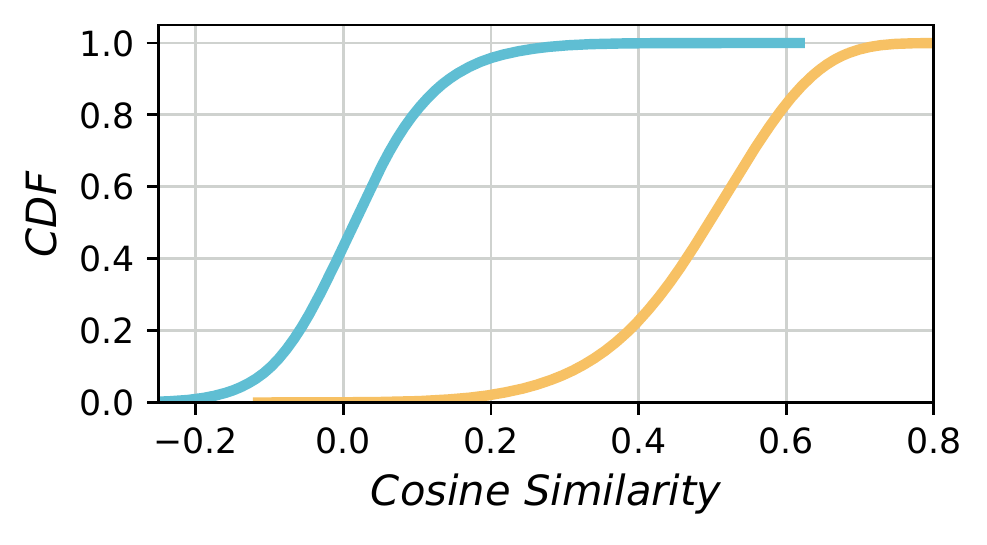}
    \label{fig:mvc_vqvc}
    }
\subfigure[VQVC+]{
    \includegraphics[height=\threesize, trim=0 0 0 0, clip]{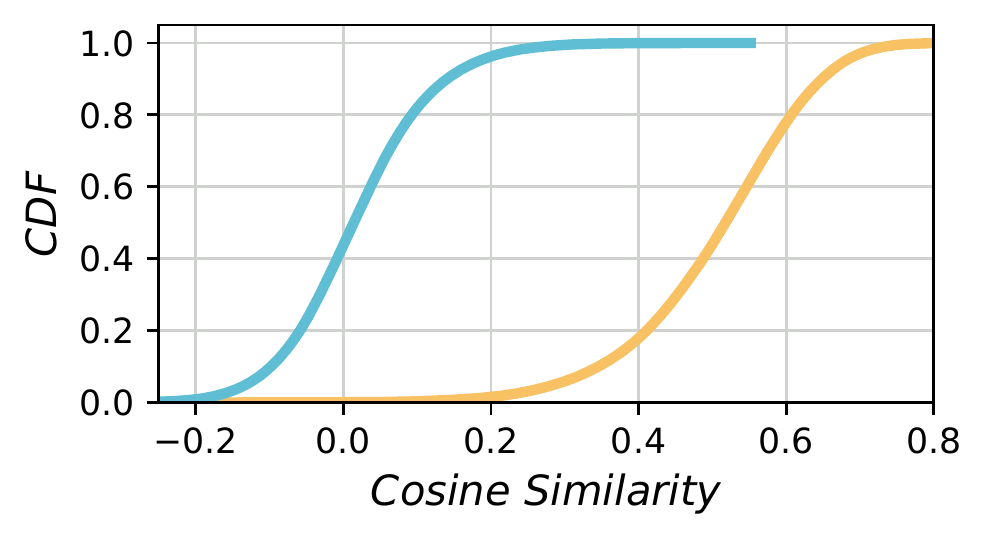}
    \label{fig:mvc_vqvcp}
    }
\subfigure[AGAIN]{
    \includegraphics[height=\threesize, trim=0 0 0 0, clip]{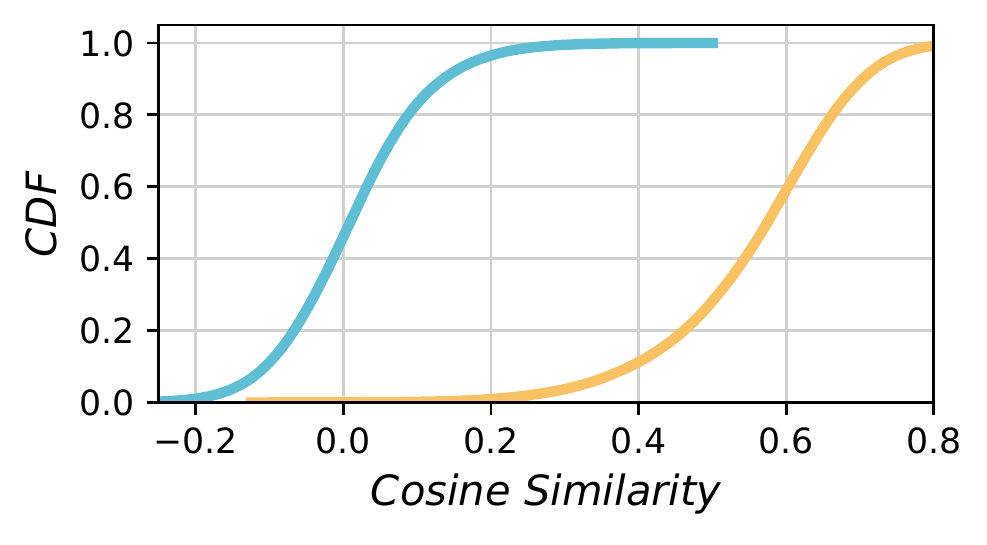}
    \label{fig:mvc_again}
    }
\subfigure[BNE]{
    \includegraphics[height=\threesize, trim=0 0 0 0, clip]{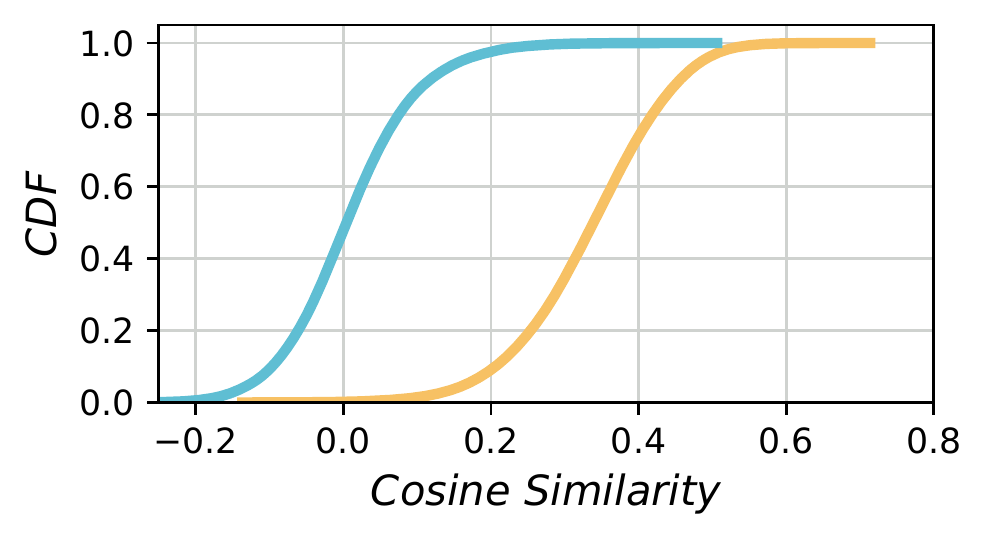}
    \label{fig:mvc_ppg}
    }

\vspace{-0.4cm}\caption{The distributions of cosine similarity scores between the restored voiceprint and the dodger or other
suspects. The voiceprints are recovered by the ensemble model of \sys. }\label{fig:dist2}
\end{figure*}


\begin{table}[ht]\centering
\resizebox{\linewidth}{!}{
\begin{threeparttable}

\caption{Voice conversion dataset.}

\small
\setlength{\tabcolsep}{1.5mm}{
\begin{tabular}{@{}llllrr@{}}
\toprule
\textbf{Method}        & \textbf{Dataset}\tnote{\textdagger} & \textbf{Alias}            & \textbf{Lang.}        & \multicolumn{1}{l}{\textbf{\#Speaker}} & \multicolumn{1}{l}{\textbf{\#Sample}} \\ \midrule

\multirow{6}{*}{VQVC}  & \emph{train-clean-100}  & \multirow{5}{*}{VQ-Train} & \multirow{5}{*}{English} & 251                                    & 62,750                               \\
                       & \emph{train-clean-360}  &                           &                          & 921                                    & 276,300                              \\
                       & \emph{train-other-500}  &                           &                          & 1,166                                  & 349,800                              \\ 
                       & VoxCeleb1        &                           &                          & 1,251                                  & 375,300                              \\ 
                       & VoxCeleb2        &                           &                          & 5,994                                  & 599,400                              \\ \cmidrule(l){2-6}
                       & \emph{test-clean}       & VQ-Test                   & English                  & 40                                     & 31,200                               \\ \midrule
\multirow{6}{*}{VQVC+} & \emph{train-clean-100}  & \multirow{5}{*}{V+-Train} & \multirow{5}{*}{English} & 251                                    & 62,750                               \\
                       & \emph{train-clean-360}  &                           &                          & 921                                    & 276,300                              \\
                       & \emph{train-other-500}  &                           &                          & 1,166                                  & 349,800                              \\ 
                       & VoxCeleb1        &                           &                          & 1,251                                  & 375,300                              \\ 
                       & VoxCeleb2        &                           &                          & 5,994                                  & 599,400                              \\ \cmidrule(l){2-6}
                       & \emph{test-clean}       & V+-Test                   & English                  & 40                                     & 31,200                               \\ \midrule
\multirow{6}{*}{AGAIN} & \emph{train-clean-100}  & \multirow{5}{*}{AG-Train} & \multirow{5}{*}{English} & 251                                    & 62,750                               \\
                       & \emph{train-clean-360}  &                           &                          & 921                                    & 276,300                              \\
                       & \emph{train-other-500}  &                           &                          & 1,166                                  & 349,800                              \\
                       & VoxCeleb1        &                           &                          & 1,251                                  & 375,300                              \\ 
                       & VoxCeleb2        &                           &                          & 5,994                                  & 599,400                              \\ \cmidrule(l){2-6}
                       & \emph{test-clean}       & AG-Test                   & English                  & 40                                     & 31,200                               \\ \midrule
                       \multirow{6}{*}{BNE}   & \emph{train-clean-100}  & \multirow{5}{*}{BN-Train} & \multirow{5}{*}{English} & 251                                    & 62,750                               \\
                       & \emph{train-clean-360}  &                           &                          & 921                                    & 276,300                              \\
                       & \emph{train-other-500}  &                           &                          & 1,166                                  & 349,800                              \\ 
                       & VoxCeleb1        &                           &                          & 1,251                                  & 375,300                              \\ 
                       & VoxCeleb2        &                           &                          & 5,994                                  & 599,400                              \\ \cmidrule(l){2-6}
                       & \emph{test-clean}       & BN-Test                   & English                  & 40                                     & 31,200                               \\ \midrule
 \multirow{3}{*}{BNE}  & MLS German       & BN-GE                     & German                   & 30                                     & 17,400                                \\
                       & MLS French       & BN-FR                     & French                   & 18                                     & 6,120                                \\
                       & MLS Spanish      & BN-SP                     & Spanish                  & 20                                     & 7,600                                \\ \bottomrule
\end{tabular}}

\begin{tablenotes}[flushleft]
\footnotesize
\item[\textdagger] \textit{train-clean-100}, \textit{train-clean-360} and \textit{train-other-500} are training sets in LibriSpeech \cite{Panayotov2015librispeech}. We use the test sets in German, French, and Spanish of Multilingual LibriSpeech (MLS) \cite{pratap2020mls} for evaluation. 
\end{tablenotes}

\label{tab:dataset}
\end{threeparttable}}
\end{table}
\raggedbottom
\begin{table}[ht]\centering
\resizebox{\linewidth}{!}{
\begin{threeparttable}
\setlength{\abovecaptionskip}{5pt}%
\setlength{\belowcaptionskip}{0pt}%
\caption{Generalization to unseen languages.}
\small

\setlength{\tabcolsep}{4.5mm}{
\begin{tabular}{@{}rrrrr@{}}
\toprule
\multicolumn{1}{r}{Metrics\tnote{\textdagger}}  & {English}        & {German}           & {French}            & {Spanish}                       \\ \midrule
 {EER ($\downarrow$)}                           & 3.78\%      & 6.08\%          & 5.78\%           & 3.68\%                     \\ \cmidrule(l){1-5}
 {Top-1~~ACC ($\uparrow$)}                      & 96.12\%     & 80.40\%         & 64.71\%          & 92.11\%                     \\
 {Top-5~~ACC ($\uparrow$)}                      & 99.97\%     & 97.30\%         & 97.22\%          & 100.0\%                     \\
 {Top-10~ACC ($\uparrow$)}                      & 100.0\%    & 99.14\%         & 99.67\%          & 100.0\%                     \\ \bottomrule
\end{tabular}}

\begin{tablenotes}[flushleft]
\footnotesize
\item[\textdagger] There are 40, 30, 18, and 20 speakers in \textit{BN-Test}, \textit{BN-GE}, \textit{BN-FR}, and \textit{BN-SP} respectively. $\uparrow$ indicates that the value is the higher, the better, and $\downarrow$ indicates that the value is the lower, the better.
\end{tablenotes}

\label{tab:language}
\end{threeparttable}}
\end{table}

\begin{table}[ht]\centering
\resizebox{\linewidth}{!}{
\begin{threeparttable}
\setlength{\abovecaptionskip}{5pt}%
\setlength{\belowcaptionskip}{2pt}%
\caption{Trained on the multi-VC dataset.}
\small

\setlength{\tabcolsep}{4.5mm}{
\begin{tabular}{@{}rrrrr@{}}
\toprule
\multicolumn{1}{r}{Metrics\tnote{\textdagger}}  & {VQVC}        & {VQVC+}           & {AGAIN}            & {BNE}                       \\ \midrule
 {EER ($\downarrow$)}                           & 2.94\%        & 2.22\%         & 1.65\%          & 4.66\%                   \\ \cmidrule(l){1-5}
 {Top-1~~ACC ($\uparrow$)}                      & 99.84\%       & 99.94\%        & 99.97\%         & 90.22\%                    \\
 {Top-5~~ACC ($\uparrow$)}                      & 100.0\%       & 100.0\%        & 100.0\%         & 99.55\%                    \\
 {Top-10~ACC ($\uparrow$)}                      & 100.0\%       & 100.0\%        & 100.0\%         & 100.0\%                    \\ \bottomrule
\end{tabular}}

 \begin{tablenotes}[flushleft]
 \footnotesize
\item[\textdagger] $\uparrow$ indicates that the value is the higher, the better, and $\downarrow$ indicates that the value is the lower, the better. 
 \end{tablenotes}
\label{tab:mvc}
\end{threeparttable}}
\end{table}

\begin{table}[ht]\centering
\resizebox{\linewidth}{!}{
\begin{threeparttable}
\caption{The detailed implementation of \sys.}

\small
\setlength{\tabcolsep}{3mm}{
\begin{tabular}{@{}p{45pt}llr@{}}
\toprule
\textbf{Module}                                                                           & \textbf{Block}                                                 & \textbf{Output size}                                                              & \textbf{\#Params}                                                  \\ \midrule
\multirow{6}{*}[-5pt]{\textbf{\begin{tabular}[c]{@{}l@{}}Feature\\ extraction\end{tabular}}}    & Filter bank                                                    & $(80, T)$                                                                  & -                                                                  \\
                                                                                                & Conv1D+ReLU+BN                                                 & $(1024, T)$                                                                & 412,672                                                            \\
                                                                                                & SE-Res2Block                                                   & $(1024, T)$                                                                & 2,713,344                                                          \\
                                                                                                & SE-Res2Block                                                   & $(1024, T)$                                                                & 2,713,344                                                          \\
                                                                                                & SE-Res2Block                                                   & $( 1024, T)$                                                                & 2,713,344                                                          \\ \cmidrule(l){2-4} 
                                                                                                & Block summary                                                 & \begin{tabular}[c]{@{}l@{}}$\mathbf{M}: (3072, T_M)$\\ $\mathbf{N}: ( 3072, {T_N}$)\end{tabular}            & 8,552,704                            \\ \midrule
\multirow{2}{*}[-2pt]{\textbf{\begin{tabular}[c]{@{}l@{}}Differential\\ rectification\end{tabular}}} & \multirow{2}{*}{\begin{tabular}[c]{@{}l@{}}Res-Orth. Block\end{tabular} }       & \multirow{2}{*}{$(3072, {T_M})$}                                                   & \multirow{2}{*}{9,446,400}                            \\ \\
                                                                                                                       \midrule
\multirow{3}{*}[-2pt]{\textbf{\begin{tabular}[c]{@{}l@{}}Dimension\\ normalization\end{tabular}}}          & Conv1D+ReLU+BN                                                 & $(3072, {T_M})  $                                                 & 9,446,400                                                          \\ 
                                                                                                & ASP+BN                                                         & $(6144, {T_M})$                                                   & 1,588,608                                                          \\
                                                                                                & FC+BN                                                          & $(192, 1)$                                                                 & 1,179,840                                                                                                              \\ \midrule
\multirow{2}{*}[-2pt]{\textbf{\begin{tabular}[c]{@{}l@{}}Voiceprint\\ enhancement\end{tabular}}}                                                                            & \multirow{2}{*}{AAM-Softmax}                                                    & \multirow{2}{*}{(9583, 1)}                                                                & \multirow{2}{*}{1,839,936}                                                   \\       \\ \midrule
\multirow{2}{*}{\textbf{\begin{tabular}[c]{@{}l@{}}Model\\ summary\end{tabular}}}               & \multicolumn{3}{c}{\multirow{2}{*}{\begin{tabular}[c]{@{}c@{}}\textbf{Input}: (1, $t_m$)$\times$(1, $t_n$) $\rightarrow$ \textbf{Output}: (9583, 1)\\\textbf{Total params}: 32,053,888\end{tabular}}}                                                                               \\
                                                                                                &                                                                                                                                    \\\bottomrule
\end{tabular}}

\begin{tablenotes}[flushleft]
\item[] \vspace{-1pt}\hspace{-2pt}\footnotesize (i) $\mathbf{M}$ and $\mathbf{N}$ are defined in \S\ref{subsec:featureextraction}. (ii) We allow the input VC audios and evidence audios to have arbitrary lengths, so ${t_m}$ and ${t_n}$ do not have to be the same.
\end{tablenotes}

\label{tab:model}
\end{threeparttable}}
\end{table}
\begin{table}[ht]\centering
\resizebox{\linewidth}{!}{
\begin{threeparttable}
\setlength{\abovecaptionskip}{5pt}%
\setlength{\belowcaptionskip}{0pt}%

\caption{Performance of voiceprint restoration on intra- and inter-gender VC.}

\small

\setlength{\tabcolsep}{3mm}{
\begin{tabular}{@{}p{25pt}rrrrr@{}}
\toprule
\textbf{Method}                     &\multicolumn{1}{r}{Metrics\tnote{\textdagger}}           & {M$\rightarrow$M}        & {F$\rightarrow$F}           & {F$\rightarrow$M}                      & {M$\rightarrow$F}                  \\ \midrule
\multirow{4}{*}[-4pt]{\textbf{VQVC}}  & {EER ($\downarrow$)}                & 3.07\%            & 4.55\%            & 5.13\%                       & 3.29\%                                              \\ \cmidrule(l){2-6}
                                & {Top-1~~ACC ($\uparrow$)}           & 94.87\%           & 92.37\%           & 90.88\%                      & 93.38\%                                              \\
                                & {Top-5~~ACC ($\uparrow$)}           & 99.08\%           & 97.76\%           & 98.00\%                      & 98.75\%                                              \\
                                & {Top-10~ACC ($\uparrow$)}           & 99.61\%           & 98.95\%           & 99.00\%                      & 99.50\%                                              \\ \bottomrule
\multirow{4}{*}[-4pt]{\textbf{VQVC+}} & {EER ($\downarrow$)}                & 3.25\%            & 3.83\%            & 3.93\%                       & 3.25\%                                     \\ \cmidrule(l){2-6}
                                & {Top-1~~ACC ($\uparrow$)}           & 95.38\%           & 93.82\%           & 93.63\%                      & 95.38\%                                     \\
                                & {Top-5~~ACC ($\uparrow$)}           & 99.25\%           & 98.95\%           & 99.13\%                      & 99.25\%                                     \\
                                & {Top-10~ACC ($\uparrow$)}           & 99.88\%           & 99.47\%           & 99.75\%                      & 99.88\%                                     \\ \bottomrule
\multirow{4}{*}[-4pt]{\textbf{AGAIN}} & {EER ($\downarrow$)}                & 1.66\%            & 2.07\%            & 2.15\%                       & 1.78\%                                     \\ \cmidrule(l){2-6}
                                & {Top-1~~ACC ($\uparrow$)}           & 99.61\%           & 98.42\%           & 99.25\%                      & 99.50\%                                     \\
                                & {Top-5~~ACC ($\uparrow$)}           & 99.87\%           & 100.0\%           & 99.75\%                      & 99.75\%                                     \\
                                & {Top-10~ACC ($\uparrow$)}           & 100.0\%           & 100.0\%           & 100.0\%                      & 100.0\%                                     \\ \bottomrule
\multirow{4}{*}[-4pt]{\textbf{BNE}}   & {EER ($\downarrow$)}                & 2.78\%            & 4.43\%            & 4.39\%                       & 2.88\%                                      \\ \cmidrule(l){2-6}
                                & {Top-1~~ACC ($\uparrow$)}           & 95.26\%           & 95.53\%           & 94.00\%                      & 93.00\%                                     \\
                                & {Top-5~~ACC ($\uparrow$)}           & 99.87\%           & 99.74\%           & 99.50\%                      & 99.75\%                                     \\
                                & {Top-10~ACC ($\uparrow$)}           & 100.0\%           & 100.0\%           & 100.0\%                      & 100.0\%                                     \\ \bottomrule
\end{tabular}}

\begin{tablenotes}[flushleft]
\footnotesize
\item[\textdagger] $\uparrow$ indicates that the value is the higher, the better, and $\downarrow$ indicates that the value is the lower, the better.

\end{tablenotes}

\label{tab:gender}
\end{threeparttable}}
\end{table}

\begin{table*}[ht]\centering
\resizebox{.6\linewidth}{!}{
\begin{threeparttable}

\setlength{\abovecaptionskip}{5pt}%
\setlength{\belowcaptionskip}{0pt}%

\caption{Over-the-telephony robustness of the ensemble model.}

\footnotesize
\setlength{\tabcolsep}{2mm}{
\begin{tabular}{@{}p{25pt}rrrrrrr@{}}
\toprule
\textbf{Method}                 &\multicolumn{1}{r}{Metrics}           & {$\mu$-law}        & {A-law}           & {GSM\tnote{\textdaggerdbl}}                      & {AMR\tnote{\textdaggerdbl}}          & {8kHz\tnote{\textdaggerdbl}}                      & {4kHz\tnote{\textdaggerdbl}}        \\ \midrule
\multirow{4}{*}[-4pt]{\textbf{VQVC}}  & {EER ($\downarrow$)}                & 8.09\%            & 8.14\%             & 9.67\%                      & 12.88\%          & {8.20\% }                     & {20.73\%}                           \\ \cmidrule(l){2-8}
                                & {Top-1~~ACC ($\uparrow$)}           & 69.14\%           & 71.89\%            & 59.36\%                     & 49.42\%          & {69.30\%}                     & {40.61\%}                           \\
                                & {Top-5~~ACC ($\uparrow$)}           & 96.31\%           & 96.64\%            & 93.69\%                     & 87.79\%          & {95.71\%}                     & {79.33\%}                           \\
                                & {Top-10~ACC ($\uparrow$)}           & 99.17\%           & 99.23\%            & 98.40\%                     & 96.41\%          & {98.65\%}                     & {91.70\%}                           \\ \bottomrule
\multirow{4}{*}[-4pt]{\textbf{VQVC+}} & {EER ($\downarrow$)}                & 6.07\%            & 5.87\%             & 7.18\%                      & 11.52\%          & {6.08\% }                     & {21.60\%}                           \\ \cmidrule(l){2-8}
                                & {Top-1~~ACC ($\uparrow$)}           & 84.33\%           & 83.69\%            & 70.42\%                     & 59.97\%          & {78.08\%}                     & {50.29\%}                           \\
                                & {Top-5~~ACC ($\uparrow$)}           & 98.69\%           & 98.27\%            & 96.25\%                     & 87.76\%          & {97.95\%}                     & {80.71\%}                           \\
                                & {Top-10~ACC ($\uparrow$)}           & 99.78\%           & 99.74\%            & 99.17\%                     & 94.71\%          & {99.68\%}                     & {91.73\%}                           \\ \bottomrule
\multirow{4}{*}[-4pt]{\textbf{AGAIN}} & {EER ($\downarrow$)}                & 2.98\%            & 2.97\%             & 3.71\%                      & 4.88\%           & {3.10\% }                     & {6.83\% }                           \\ \cmidrule(l){2-8}
                                & {Top-1~~ACC ($\uparrow$)}           & 98.88\%           & 98.65\%            & 96.92\%                     & 93.14\%          & {98.43\%}                     & {90.03\%}                           \\
                                & {Top-5~~ACC ($\uparrow$)}           & 100.0\%           & 99.97\%            & 99.94\%                     & 99.65\%          & {100.0\%}                     & {99.87\%}                           \\
                                & {Top-10~ACC ($\uparrow$)}           & 100.0\%           & 100.0\%            & 100.0\%                     & 99.90\%          & {100.0\%}                     & {99.97\%}                           \\ \bottomrule
\multirow{4}{*}[-4pt]{\textbf{BNE}}   & {EER ($\downarrow$)}                & 5.43\%            & 5.52\%             & 6.22\%                      & 10.59\%          & {5.42\% }                     & {11.65\%}                           \\ \cmidrule(l){2-8}
                                & {Top-1~~ACC ($\uparrow$)}           & 84.52\%           & 84.68\%            & 80.58\%                     & 62.72\%          & {85.67\%}                     & {63.46\%}                           \\
                                & {Top-5~~ACC ($\uparrow$)}           & 98.85\%           & 98.97\%            & 98.01\%                     & 87.69\%          & {98.43\%}                     & {90.35\%}                           \\
                                & {Top-10~ACC ($\uparrow$)}           & 99.87\%           & 99.78\%            & 99.68\%                     & 94.65\%          & {99.90\%}                     & {96.89\%}                           \\ \bottomrule
\end{tabular}}             

\begin{tablenotes}[flushleft]
\footnotesize

\item[\textdaggerdbl] Short for GSM-FR, AMR-NB, 8kHz subsampling, and 4kHz subsampling.

\end{tablenotes}

\label{tab:robust_mvc}
\end{threeparttable}}
\end{table*}






\end{document}